\documentclass{aa}

\usepackage{subcaption}
\usepackage{float}
\usepackage{appendix}
\usepackage{graphicx}
\usepackage{threeparttable}
\usepackage{multirow}
\usepackage{txfonts}
\usepackage[%
    colorlinks=true,
    citecolor=blue,
    linkcolor=blue
]{hyperref}
\usepackage{caption}
\DeclareCaptionLabelFormat{boldlabel}{\textbf{#1~#2}}
\titlerunning{A JVLA, GMRT, and {\it XMM} study of Abell 795}
\authorrunning{N. Rotella et al.}

\begin{document}

   \title{A JVLA, GMRT, and {\it XMM} study of Abell 795: Large-scale sloshing and a candidate radio phoenix}

   \author{N. Rotella\inst{1},
          F. Ubertosi\inst{1,}\inst{2},
          M. Gitti\inst{1,}\inst{2}, M. Rossetti\inst{3}, F. Gastaldello\inst{3}, G. W. Pratt\inst{4}, F. Brighenti\inst{1,5}, E. Torresi\inst{6} \and P. Grandi\inst{6}}

   \institute{Dipartimento di Fisica e Astronomia, Università di Bologna, Via Gobetti 93/2, 40129 Bologna, Italy\\
              \email{nicolo.rotella@studio.unibo.it}
         \and
             Istituto Nazionale di Astrofisica – Istituto di Radioastronomia (IRA), Via Gobetti 101, 40129 Bologna, Italy
         \and
            Istituto Nazionale di Astrofisica - Istituto di Astrofisica Spaziale e Fisica cosmica (IASF), Via A. Corti 12, I-20133 Milano, Italy
        \and
            Université Paris-Saclay, Université Paris Cité, CEA, CNRS, AIM de Paris-Saclay, 91191 Gif-sur-Yvette, France
        \and
            University of California Observatories/Lick Observatory, Department of Astronomy and Astrophysics, Santa Cruz, CA 95064, USA
        \and
            INAF-Osservatorio di Astrofisica e Fisica dello Spazio, Via Gobetti 101, I-40129, Bologna, Italy
             }

   \date{Received January 21, 2025; accepted March 26, 2025}

\abstract
    {}
   {The galaxy cluster Abell 795 (z = 0.1374) is known from previous works for the presence of extended  ($\approx$ 200 kpc) radio emission with a steep spectral index of an unclear origin surrounding the brightest cluster galaxy (BCG), and for the sloshing signatures visible in {\it Chandra} observations of its cool core. Our purpose is to investigate the nature of the extended radio emission and its possible link with the dynamical state of the intracluster medium (ICM) on large scales ($\approx$ 1 Mpc).}
   {We used new JVLA 1.5 GHz, as well as archival GMRT 325 MHz and {\it XMM-Newton} X-ray observations to study the cluster with a thermal and nonthermal approach. }
   {From the {\it XMM} surface brightness analysis, we detected an azimuthally asymmetric excess reaching around 650 kpc from the center of Abell 795. The excess appears to follow the sloshing spiral previously detected, but with the existing {\it XMM} data it is not possible to confirm its classification as a large-radius cold front in Abell 795. Furthermore, the X-ray data allowed us to detect the hot gas from a previously unknown galaxy group at a projected distance of $\approx$ 7.4$\arcmin$ (1 Mpc) northwest of Abell 795. Its surface brightness radial profile is well-fitted with a $\beta$ model of slope $\beta = 0.52\pm0.17$, and the spectral analysis reveals a thermal plasma of temperature $kT = 1.08\pm0.08$~keV and metallicity Z $=0.13\pm0.06$~Z$_{\odot}$. We discuss the possibility that this galaxy group is the perturber that initiated sloshing in Abell 795, and we show that the velocity distribution of member galaxies supports the dynamically unrelaxed nature of Abell 795. The analysis of JVLA 1.5 GHz and GMRT 325 MHz images confirms the presence of extended radio emission with the largest linear size $\approx$ 200 kpc, preferentially extended toward southwest and terminating in a sub-component ("SW blob"). We measured the spectral indices between 325 MHz and 1.5 GHz, finding $\alpha_{Ext} = -2.24\pm0.13$ for the diffuse extended emission, and $\alpha_{SWb} = -2.10 \pm0.13$ for the SW blob. These ultra-steep spectral index values, coupled with the complex morphology and cospatiality with the radio-loud active galactic nuclei (AGN) present in the BCG, suggest that this extended emission could be classified as a radio phoenix, possibly arising from adiabatic compression of an ancient AGN radio lobe due to the presence of sloshing motions.}
    {}

   \maketitle
%

\section{Introduction}

In recent years, increasing attention has been devoted to the study of diffuse radio sources in galaxy clusters. Indeed, with the advent of sensitive and high spatial resolution radio telescopes -- such as the Low Frequency Array (LOFAR),  the Karl G. Jansky Very Large Array (JVLA), the upgraded Giant Metrewave Radio Telescope (uGMRT), and MeerKAT -- the window onto the Universe at low frequencies has revealed a growing number of such sources (e.g., \citealt{giova}, \citealt{ferett}, \citealt{vanWeeren}, \citealt{Know}). Several classification schemes have been proposed to better understand the diverse category of diffuse sources. Here we consider one of the most recent ones by \cite{vanWeeren}, which includes radio halos (giant and mini), cluster radio shocks, and revived active galactic nuclei (AGN) fossil plasma sources (e.g., phoenices, gently re-energized tail).\\
The origin of diffuse radio emission in galaxy clusters remains an open question. The relativistic electron populations that should be responsible for the extended synchrotron emission in clusters typically have $t_{age} \leq 10^{8}$ yr, which implies a diffusion length scale on the order of $\approx10^{2}$ pc  \citep[e.g.,][]{diffusion_length}. This suggests that extended emission on scale of hundreds of kiloparsec cannot be attributed to relativistic electrons accelerated at a single location in the intracluster medium (ICM), a challenge known as the slow diffusion problem. To explain this, the electrons must either undergo reacceleration or be produced in situ \citep{in-situ}. While progress has been made in understanding the mechanisms behind the formation of radio halos (e.g., \citealt{gittiold}, \citealt{brune}, \citealt{cuciti21}, \citealt{cuciti23}) and cluster radio shocks (e.g., \citealt{vanWee2011}, \citealt{molnar}, \citealt{bott20}, \citealt{jones23}), the origin of revived AGN plasma sources is still less clear. This is primarily due to the limited number of known examples, but also to the presence of the radio AGNs in the brightest cluster galaxy (BCG), which can further complicate their study.\\
Active galactic nuclei in galaxy clusters are a natural source of relativistic electrons that age over time due to radiative losses. Therefore, they may provide the seed particles for the formation of diffuse radio emission (e.g., \citealt{greet}). In particular, it has been proposed that phoenices trace old radio plasma emitted from past episodes of AGNs activity \citep[e.g.,][]{vanWeeren}. Shock waves moving in the ICM can compress the seed relativistic electrons, boosting their momentum and, in the presence of magnetic fields, producing synchrotron emission characterized by a steep spectral index through adiabatic compression \citep{adiabatic}. However, direct observational evidence for a connection between shock waves and phoenices is still missing; therefore their formation scenario remains somewhat uncertain. The spectral index\footnote{The spectral index $\alpha$ is defined as S$_{\nu} \propto \nu^{\alpha}$ (where $\nu$ is the frequency
and S$_{\nu}$ is the flux density at the frequency $\nu$).} distribution across these sources is irregular without clear common trends \citep{Cohen} and with global values $\alpha \leq -1.5$ (e.g., \citealt{Cohen, degasp1}). Only a few polarization studies have been performed so far of these sources \citep{slee}, finding low polarization levels (2-35 $\%$).
\begin{figure*}[ht!]
    \centering
    \sidecaption\includegraphics[width=0.72\linewidth]{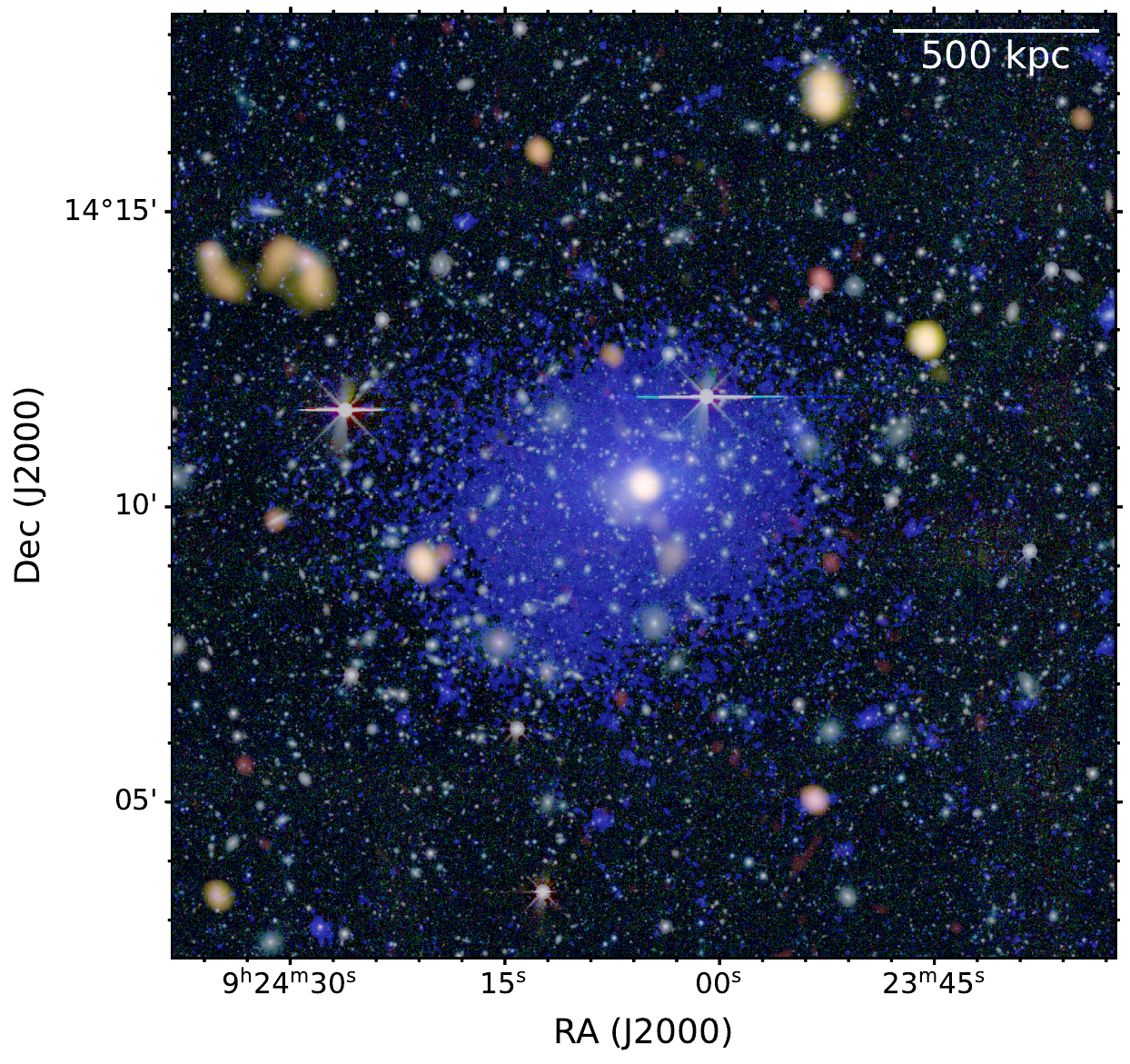}
    \caption{Composite optical, radio, and X-ray image of A795. The optical image in the background is from the DESI survey. Blue corresponds to the {\it XMM}-EPIC (0.5 -- 2 keV) exposure-corrected image, while yellow and red correspond to the GMRT 325 (restoring beam of 15\arcsec x 15\arcsec) MHz and JVLA 1.5 GHz (restoring beam of 12.8\arcsec x 10.4\arcsec) respectively.}
    \label{A795multi}
\end{figure*}
The elongated and filamentary morphologies are the most common for phoenices (e.g., \citealt{morpho}, \citealt{slee}), with sizes $\leq 300 - 400$ kpc. Recently, a new formation scenario has been proposed for phoenices, in which the relativistic plasma expelled by AGNs may interact with the gas which is sloshing in the cluster gravitational potential well, as observed in Abell 2657 \citep{botteon}. This scenario, in which sloshing motions are able to disrupt AGN-plasma lobes and form phoenices was studied in simulations done by \cite{zuhone1}.\\
Cold fronts were first discovered in galaxy clusters by \citet{cold1}, where they were interpreted as contact discontinuities in pressure equilibrium, distinct from shocks. These features are characterized by a sharp temperature jump across the front, with lower temperatures in regions of a higher density and higher temperatures in regions of a lower density, while maintaining nearly constant pressure. Subsequent studies revealed that cold fronts are ubiquitous in many galaxy clusters, including those considered to be dynamically relaxed (e.g., \citealt{reviewcold}, \citealt{ghizza}).
A leading interpretation for the origin of cold fronts in dynamically relaxed clusters involves the sloshing mechanism, where the ICM undergoes oscillations within the gravitational potential well of the cluster, triggered by a minor merger with a large impact parameter \citep{ascasibar}. These oscillations produce distinctive features in X-ray images, such as spiral-like structures or staggered arcs, depending on the viewing angle and the angular momentum generated. The typical sizes of these structures are on the order of $\approx10^{2}$ kpc \citep{cold}.
The perturber initiating the sloshing process, often a smaller cluster or galaxy group, experiences ram pressure stripping as it moves through the ICM. This interaction can lead to the removal of gas from the perturber, forming an extended gas structure along its motion, and in later stages, possibly resulting in the complete depletion of its gas content \citep{ascasibar}. Additionally, cold fronts have been associated with mini radio halos, particularly in cool-core clusters, linking sloshing-induced turbulence to the reacceleration of relativistic electrons in the cluster core \citep{mazzotta}.\\
The galaxy cluster Abell 795 \citep[hereafter A795, z = 0.1374,][]{Rines},  located at RA 09:24:05.3, Dec +14:10:21.5 (J2000), represents a relevant case of study in this context.
The first dedicated X-ray study of A795 was performed by \cite{a795}, who analyzed {\it Chandra} observations and classified it as a weak cool-core cluster. They detected sloshing signatures in the form of two cold fronts at $\approx$ 60 kpc and $\approx$ 180 kpc from the BCG, and also explored the potential feedback on the ICM from the radio BCG, which had been previously classified as a Fanaroff-Riley \citep{FR} Type 0 radio galaxy (\citealt{baldi1}, \citealt{baldi}). Furthermore, the analysis of archival GMRT data at 150 MHz at 25$\arcsec$ resolution revealed a slightly resolved, extended radio structure spanning approximately 200-300 kpc with a complex morphology. \cite{a795} proposed this extended emission as a candidate mini-halo based on its steep ($\alpha\leq-1.5$) spectral index and shape, as well as its association with sloshing, but it remained to be confirmed by deeper multifrequency radio observations.\\
Recently, \cite{newa795} analyzed archival GMRT observations at both 150 MHz (the same examined by \citealt{a795}) and 325 MHz finding extended radio emission with an ultra-steep spectral index ($\alpha$ = $-$2.71 $\pm$ 0.28) that they classified as a radio mini-halo. Furthermore, their reanalysis of archival {\it Chandra} data confirmed the spiral-shaped morphology of the ICM in A795.\\
In this work, we analyzed new observations of A795 performed with the JVLA at 1.5 GHz, as well as archival GMRT data at 325 MHz (the same studied by \citealt{newa795}), and present the first in-depth individual study of archival {\it XMM-Newton} observations of this cluster. \\
The paper is organized as follows: Sec. \ref{data} describes the data reduction procedures for the JVLA, GMRT, and {\it XMM-Newton} observations. In Sec. \ref{morphology}, we analyze the radio and X-ray morphology of A795, focusing on the extended radio emission, the sloshing features in the intracluster medium, and the identification of a candidate galaxy group. The spectral analysis of both radio and X-ray data, including the thermodynamic properties of the ICM and the candidate group, is detailed in Sec. \ref{spectral}. In Sec. \ref{discussion}, we discuss the implications of our findings, focusing on the sloshing-induced features, the possible interaction between A795 and the candidate galaxy group, and the classification of the extended radio emission as a potential radio phoenix. Finally, we summarize our conclusions in Sec. \ref{conclusions}.\\
The cosmology adopted is $H_{0} =$ 73 km/s/Mpc, $\Omega_{m}$ = 0.3, $\Omega_{\Lambda}$ = 0.7, which gives a conversion scale factor of  2.43 kpc/arcsec at the redshift of A795.

\section{Data reduction}
\label{data}
\subsection{Radio data: JVLA and GMRT} We analyzed new JVLA observations of A795 (Project code 22B-196, PI: Ubertosi) performed with the C configuration at L band (1 -- 2 GHz) for a total time of 1 hour. The source 3C 147 was used as the flux density and bandpass calibrator, while J0854+2006 was used as the phase calibrator. The data were calibrated in \textit{CASA} v.6.4.4 \citep{casa} using standard data reduction techniques for continuum calibration. We also performed two rounds of phase self-calibration and one round of phase and amplitude self-calibration, which significantly reduced noise levels across the field.\\
In addition, in this work we employed GMRT observations at 325 MHz (Observation 4715, January 14, 2010), with 4.2 hours of time on source. We processed the archival GMRT observations using the Source Peeling and Atmospheric Modeling pipeline (SPAM; \citealt{spam}). SPAM performs the standard data reduction and calibration steps (bandpass and gain calibration, direction-independent self-calibration), as well as direction-dependent self-calibration. We used the calibrator 3C~286 to set the flux density scale and determine the bandpass shape in SPAM. \\
The data were imaged in \textit{CASA} using the \texttt{tclean} task, exploring various weighting schemes and deconvolution algorithms to optimize the results. We compared \texttt{robust} = 0.5, which balances angular resolution and sensitivity, with \texttt{robust} = 2 (natural weighting), which maximizes sensitivity at the expense of angular resolution. As expected, the natural-weighted image achieved the lowest RMS noise level (e.g., at 1.5 GHz, natural weighting yields an RMS noise of 27 $\mu$Jy/beam, compared to 35 $\mu$Jy/beam for the \texttt{briggs} = 0.5 image).
We found the "mtmfs" algorithm \citep{mtmfs} to be the most effective deconvolution method, as it accounts for variations in spectral index across the field and for emission on different spatial scales.\\
The errors on JVLA and GMRT flux densities were computed using the formula
\begin{equation}
    \Delta S = \sqrt{(\sigma_{c} \cdot S)^{2} + (RMS \cdot \sqrt{n_{beam}})^{2}}\\,
\end{equation}
where $n_{beam}$ is the number of beams in the source area and $\sigma_{c}$ is the systematic uncertainty in the flux density calibration, assumed to be 5\% for JVLA \citep{condon} and 8\% for GMRT \citep{gmrt}.
\begin{figure*}[ht!]
    \centering
    \sidecaption
    \includegraphics[width=0.7\textwidth, height=0.4\textwidth]{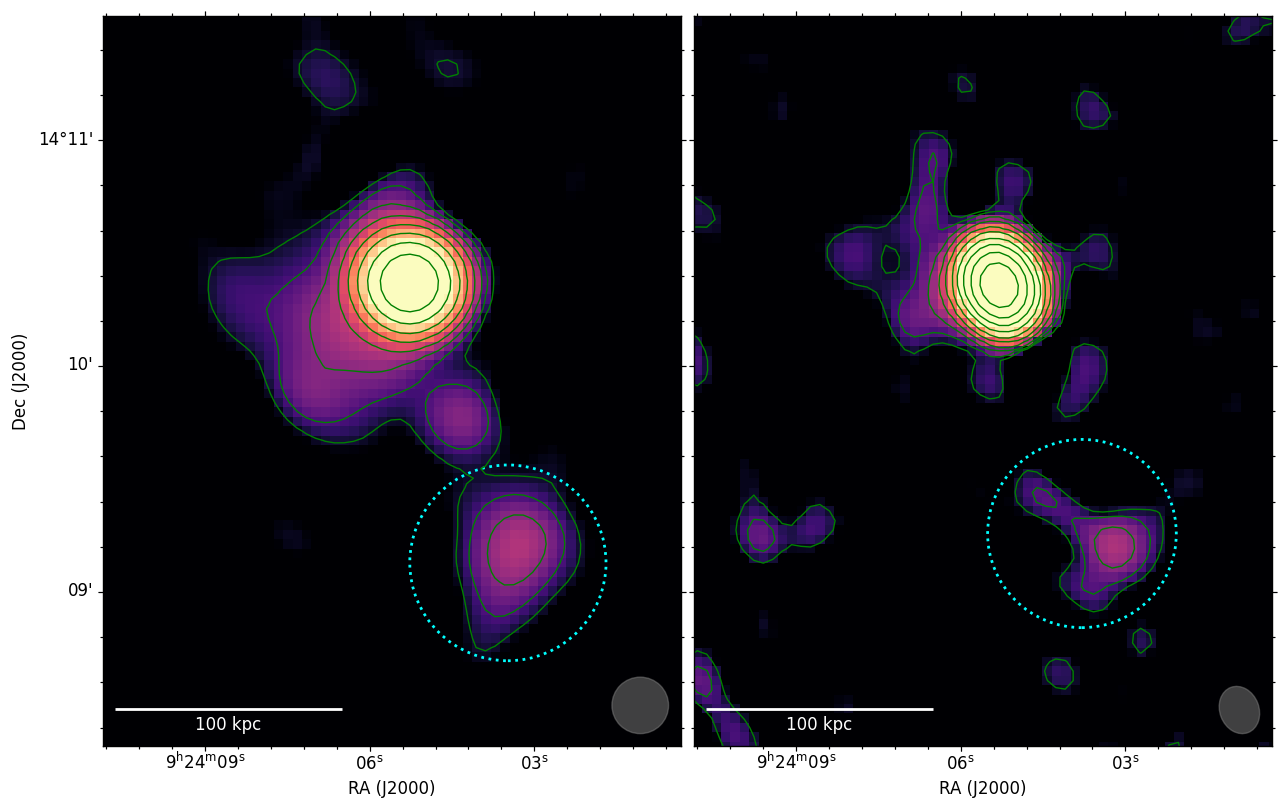}
    \caption{Radio images of the central radio source in A795 obtained with \texttt{briggs} = 0.5. \textit{Left panel}: GMRT 325 MHz image; the restoring beam of $15\arcsec \times 15\arcsec$ is shown in gray. RMS is 520 $\mu$Jy/beam. \textit{Right panel}: JVLA 1.5 GHz image; the restoring beam of $12.8\arcsec \times 10.4\arcsec$ is shown in gray. RMS is 35 $\mu$Jy/beam. In both panels, the contours are drawn at 3, 6, 12, 24, 48 $\times$ RMS and the dotted circle marks the SW blob.}
    \label{combined}
\end{figure*}
\subsection{X-ray data: {\it XMM-Newton}}
We analyzed an archival {\it XMM-Newton} data set (Observation ID: 0827031001) with an exposure time of 34 ks. The source A795 was observed with the European Photon Imaging Camera (EPIC), which consists of two MOS \citep{mos} cameras and one pn \citep{pn} camera.
The dataset was reduced using the X-COP pipeline, which was specifically designed for the analysis of X-COP observations (\citealt{pip2},  \citealt{pip3}, \citealt{pyprof}). This pipeline performs data  processing using the \textit{Extended Source Analysis Software (ESAS}, \citealt{snowden}), which is integrated into \textit{SAS} v.16.1, starting from the observation data file (ODF) through to the 2D spectral analysis results.\\
After examining the light curves with the \texttt{mos-filter} and \texttt{pn-filter} tasks, we found no soft proton flares, allowing us to use the entire observation duration. We then applied the \texttt{pn-spectra} and \texttt{mos-spectra} tasks to the event files to produce images and spectra of the source. The \texttt{pn-back} and \texttt{mos-back} tasks were used to compute the particle background images and spectra. EPIC images (0.5 -- 2 keV) were obtained by merging the pn and MOS data, and point sources were removed using the \texttt{ewavelet} task.\\
For the surface brightness analysis, we used \textit{pyproffit} \citep{pyproff}, a python package specifically designed for the morphological analysis of cluster of galaxies. Spectral fitting was performed with \textit{Xspec} v.12.13.0.

\section{Radio and X-ray morphology of A795}
\label{morphology}
In this section, we thoroughly analyze the morphology of A795 in the radio and X-ray bands. First, we present the JVLA and GMRT images of the radio sources in the field of A795. Then, we examine the X-ray surface brightness distribution of the ICM in A795. Figure \ref{A795multi} shows a composite optical, radio, and X-ray image of A795.

\subsection{The radio view}
\label{radio}
The final GMRT 325 MHz and JVLA 1.5 GHz images are presented in Fig. \ref{combined}. These images were produced with \texttt{robust} = 0.5, while those obtained with \texttt{robust} = 2 are shown in Fig. \ref{combined_natural} in the Appendix. These images display the central region of A795, where the radio AGN in the BCG coincides with the flux density peak. Consistent with previous work (\citealt{a795}, \citealt{newa795}), we confirm the detection of two extended structures: the emission surrounding the BCG and extended emission towards the southwest direction (hereafter referred to as the SW blob; see the dashed circle in Fig. \ref{combined}).
\begin{table*}
\caption{Flux densities and largest linear size (LLS) of the radio source components at the center of A795.}
    \centering
    \renewcommand{\arraystretch}{1.5}
    \begin{tabular}{c  c  c  c  c }
        \hline
         & \multicolumn{1}{c}{$S_{1.5 GHz}$ (mJy)} & $S_{325 MHz}$ (mJy) & $LLS_{1.5 GHz}$ (kpc)[arcsec] & $LLS_{325 MHz}$ (kpc)[arcsec]\\
        \hline
        BCG & 96.7 $\pm$ 5.0 & 539.2 $\pm$ 38.5 & -- & --\\
        Extended em. & 3.24 $\pm$ 0.35 & 84.3 $\pm$ 4.5 & 170 [70] & 217 [89]\\
        SW blob & 1.10 $\pm$ 0.02 & 24.1 $\pm$ 2.4 & 97 [40] & 110 [45]\\
        \hline
    \end{tabular}\\

\tablefoot{Measurements are taken from the radio images at 1.5 GHz and 325 MHz with \texttt{robust} = 0.5, see Fig. \ref{combined}. The value for the extended emission is obtained by subtracting the flux density of the BCG from the total flux density (excluding the SW blob).}
    \label{fluxes}
\end{table*}
Both components appear more extended at 325 MHz than at 1.5 GHz. The emission surrounding the BCG has a largest linear size (LLS) of $\approx$ 170 kpc in the 1.5 GHz data, while it reaches $\approx$ 217 kpc in the 325 MHz image. Similarly, the SW blob, located $\approx$ 201 kpc from the cluster center, exhibits a LLS of $\approx$ 90 kpc at 1.5 GHz but extends to $\approx$ 110 kpc at 325 MHz.\\
Measuring the flux density of the extended emission requires a careful subtraction of the cospatial radio emission from the AGN in the BCG. The radio emission from the central AGN is unresolved at 325 MHz and 1.5 GHz (as expected given its classification as a FR0 radio galaxy, see \citealt{torresi}, \citealt{a795}). Therefore, to isolate the AGN emission, we produced an image of the point sources only, by restricting the uv-range during imaging to 7 -- 22 k$\lambda$. The lower bound corresponds to scales of 30$\arcsec$, which is twice the beam FWHM, and is set to avoid the inclusion of any extended emission. Images at 1.5 GHz and 325 MHz were created by imposing the above uv-range and by selecting \texttt{robust} = 0.5. The results are shown in Fig. \ref{fr0} in the Appendix. Using this method, we measured the flux densities at 325 MHz and 1.5 GHz corresponding to the unresolved core of the central AGN. Then, we obtained the flux densities of the extended emission by subtracting the point-source flux density from the total flux density measured in the images obtained from the full uv-range.
In Table \ref{fluxes} we summarize the flux density values and LLS of the region used to measure the total flux density (within 3 RMS contour) of the central extended emission, the BCG and the SW blob.

\subsection{X-ray morphological analysis}
\label{morphologyX}
\begin{figure*}[!ht]
\captionsetup[subfigure]{labelformat=empty}
  \subfloat[][]{\includegraphics[width=0.48\linewidth]{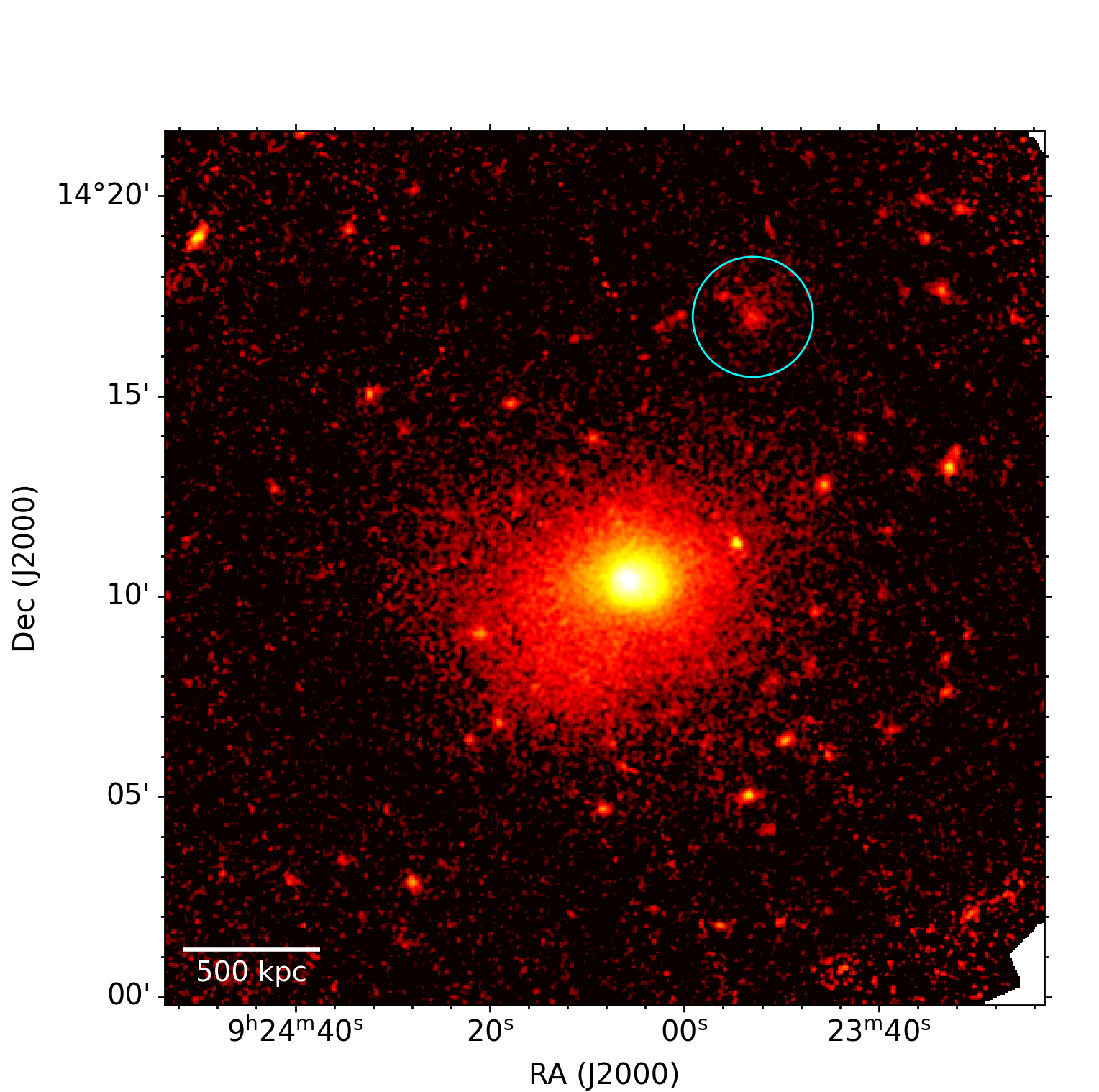}}\hfill%
  \subfloat[][]{\includegraphics[width=0.48\linewidth]{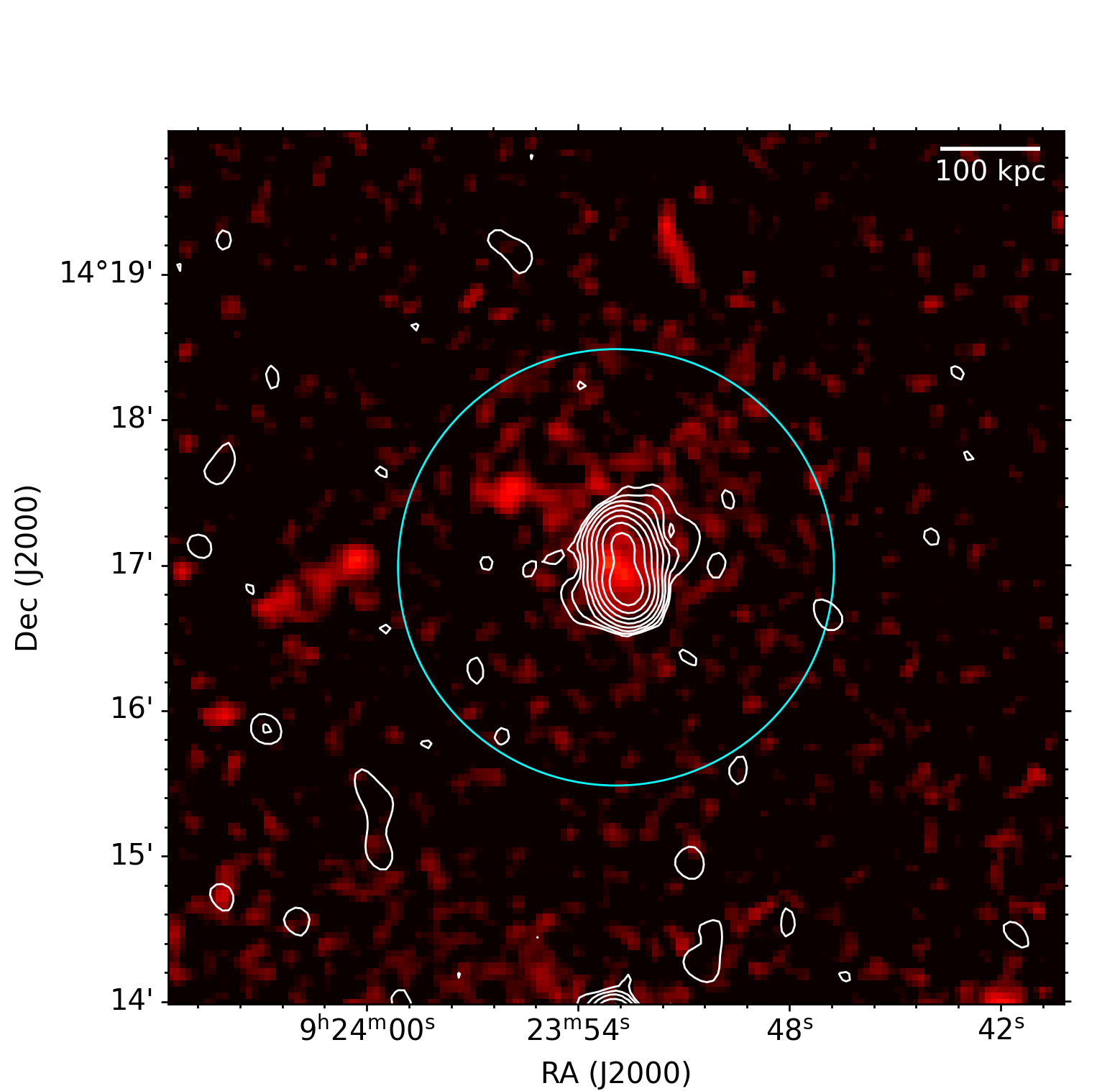}}\vspace*{-1cm}\par
  \subfloat[][]{\includegraphics[width=0.48\linewidth]{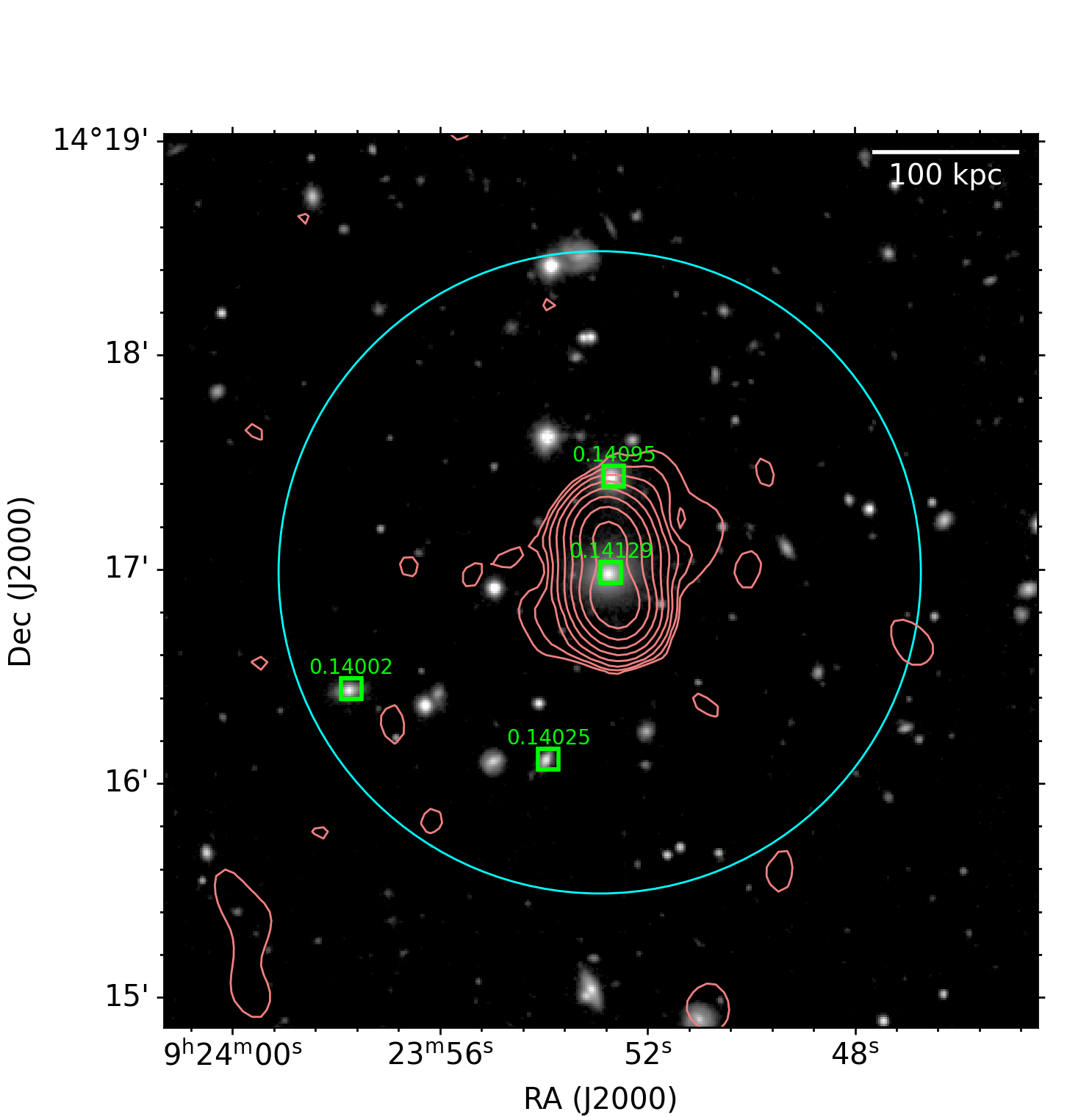}}\hfill%
  \parbox[b][.32\textheight]{.42\textwidth}%
    {\vskip-1mm{\caption{{\it XMM}-EPIC (0.5 -- 2 keV) exposure-corrected image Gaussian-smoothed with a kernel radius of 3 pixels shown on top, and below it, the optical image from the DESI survey. The cyan circle marks the position of an additional extended X-ray source (separate from the ICM of A795), see Sec. \ref{group2D}. The point sources detected are marked with white circles in left panel of Fig. \ref{spectrum}. \textit{Top left panel}: wide field view of A795. \textit{Top right panel}: zoom-in on the extended X-ray source within the cyan circle in the left panel. The white contours are from the JVLA 1.5 GHz data (same as in right panel of Fig. \ref{combined}). \textit{Bottom left panel}: optical image from the DESI survey at the location of the extended X-ray source. The green squares mark the positions of four galaxies with similar spectroscopic redshift (from \citealt{Rines}), which are reported on the image. The light red contours are from the JVLA 1.5 GHz data (same as in right panel of Fig. \ref{combined}). In each panel, the cyan circle has a radius of 1.5\arcmin, roughly corresponding to 200~kpc at the redshift of A795 ($z = 0.1374$).}\vfill\label{fluxim}}}
\end{figure*}

\begin{figure*}
    \centering
    \begin{subfigure}{.45\textwidth}
        \includegraphics[height=6.5 cm]{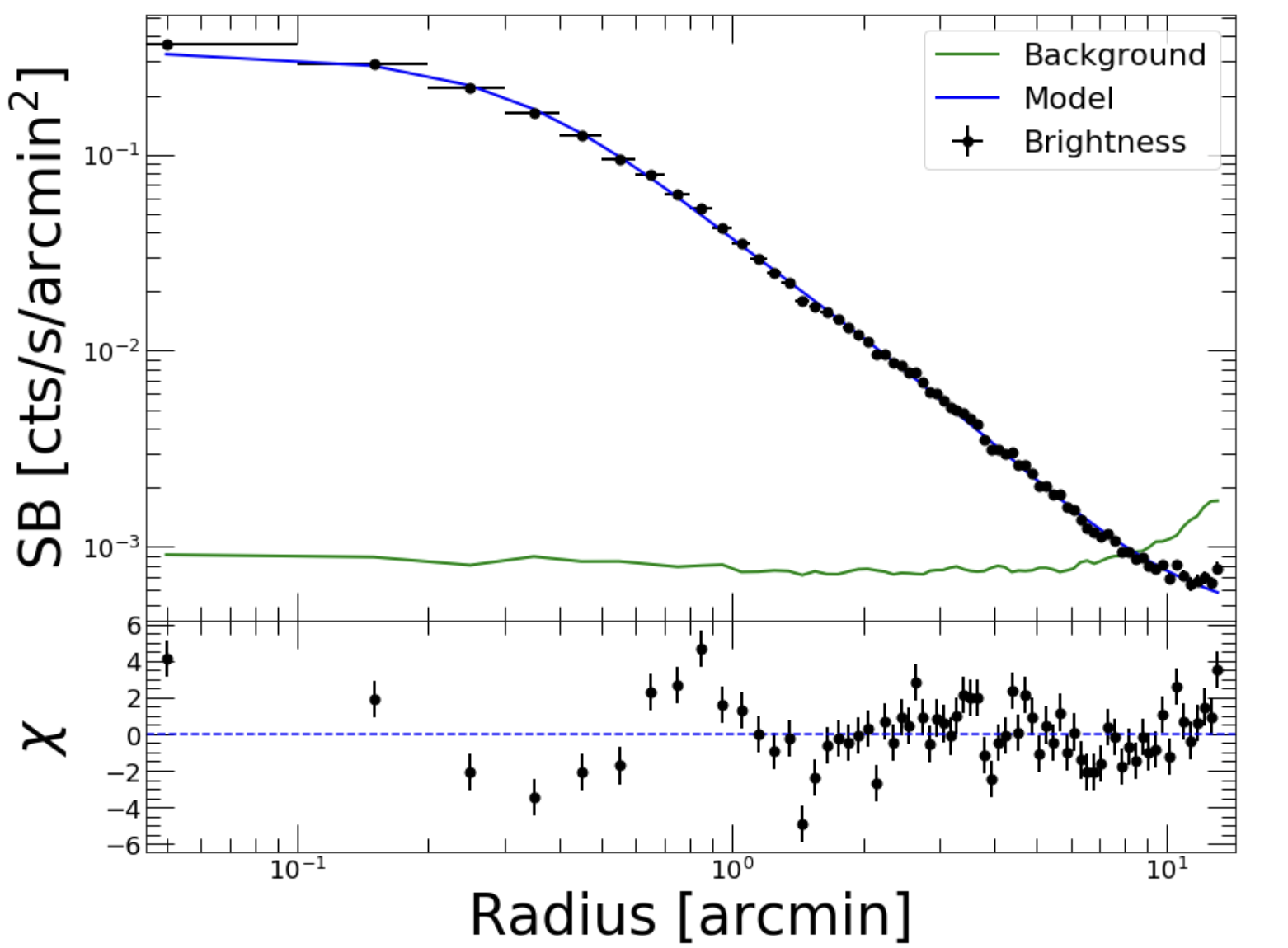}
    \end{subfigure}
    \hspace{0.1cm} %
    \begin{subfigure}{.5\textwidth}
        \includegraphics[height=6.5cm]{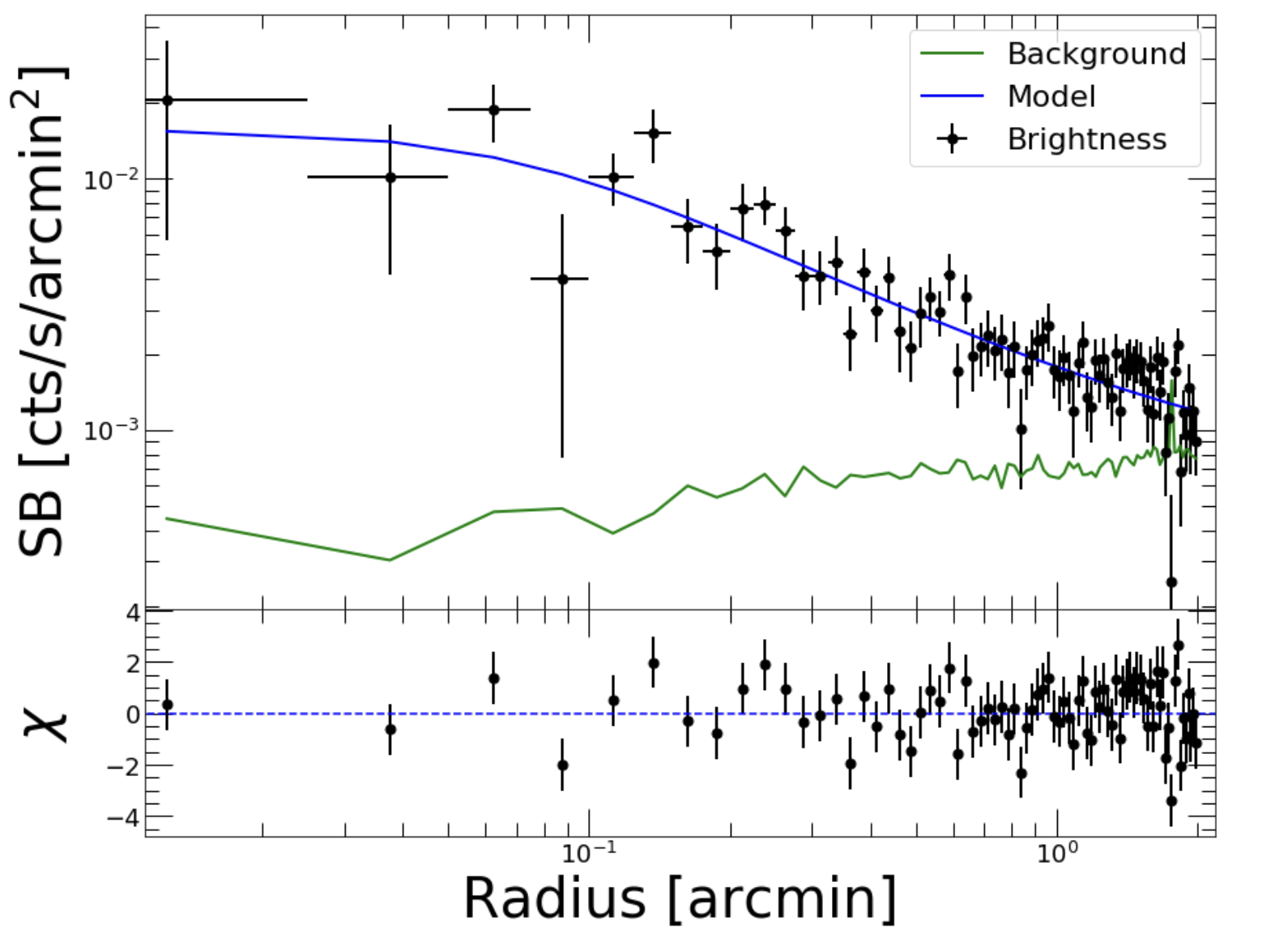}
    \end{subfigure}
    \caption{{\it XMM}-EPIC (0.5 -- 2 keV) surface brightness profiles of the galaxy cluster A795 and the candidate galaxy group. Black points are the surface brightness measurements, while the green line represents the particle background. The lower panel shows the residuals from the best-fit model. \textit{Left panel}: A795 profile fitted with a double $\beta$ model. The best-fit parameters are reported in Table \ref{beta}. \textit{Right panel}: Surface brightness profile of the extended X-ray source 7.36$\arcmin$ northwest of A795, that we interpret as a candidate galaxy group at a similar redshift of A795 (see Sec. \ref{group2D} and Fig. \ref{fluxim}), fitted with a single $\beta$ model. The best-fit parameters are reported in Table \ref{beta} (last row).}
    \label{Sb}
\end{figure*}

The left panel in Fig. \ref{fluxim} shows the exposure-corrected image of A795 in the 0.5 -- 2 keV energy band, which was Gaussian-smoothed with a kernel radius of 3 pixels. The X-ray surface brightness is centrally peaked and shows a southeast extension at $\approx$ 600 kpc from the center.
\\
In order to investigate the global surface brightness profile of A795, we extracted the profile in \textit{pyproffit} using concentric circular annuli with logarithmically spaced binning starting from a radius 6$\arcsec$, with an outer radius of 13.5$\arcmin$ (1.97 Mpc). As the center of the annuli, we selected the position of the X-ray peak (RA, Dec = 09:24:05.69, +14:10:26.09) after confirming its agreement with the X-ray peak found in the {\it Chandra} data from \cite{a795}. The {\it Chandra} data also revealed an offset of $\approx$ 7$\arcsec$ (18 kpc) between the X-ray peak and the position of the BCG (RA, Dec = 09:24:05.30, +14:10:21.51).
\begin{table*}
\caption{Best-fit results of the X-ray surface brightness profile modeling.}
    \centering
    \renewcommand{\arraystretch}{1.5}
    \begin{tabular}{c c  c  c  c  c }
        \hline
         System & Model & \multicolumn{1}{c}{$\beta$} & $r_{c,1}$ (kpc) & $r_{c,2}$ (kpc) & $\chi^{2}$ / \textit{D.O.F.} \\
        \hline
        \multirow{2}{*}{A795}& Single-$\beta$ & 0.48 $\pm$ 0.02 & 45.7 $\pm$ 5.2 & -- & 257.1 / 67 (3.84)\\
        & Double-$\beta$ & 0.59 $\pm$ 0.03 & 59.8 $\pm$ 6.7 & 317.8 $\pm$ 34.5 & 216.0 / 65 (3.32) \\
        \hline
        Galaxy group (A795-B) & Single-$\beta$ & 0.52 $\pm$ 0.17 & 28.2 $\pm$ 13.1 & -- & 54.51 / 56 (0.97)\\
        \hline
    \end{tabular}
    \tablefoot{The first two rows report the best-fit parameters of the single and double $\beta$ model for the surface brightness profile of A795. The last row shows the best-fit parameters of the single $\beta$ model fitted to the surface brightness profile of the extended X-ray source 7.36$\arcmin$ northwest of A795, that we interpret as a candidate galaxy group at a similar redshift of A795 (see Sec. \ref{group2D} and Fig. \ref{fluxim}). The profiles for A795 and the galaxy group are shown in left and right panels of Fig. \ref{Sb}, respectively.}
    \label{beta}
\end{table*}
We fitted the {\it XMM} surface brightness profile of A795 using both a spherical single $\beta$ model \citep{betamod} and a spherical double  $\beta$ model \citep{mohr} with linked $\beta$, see left panel in Fig. \ref{Sb}. The green line in the plot represents the particle background profile, which is subtracted from the source one, and the lower panel shows the residuals from the best-fit model. The results of the fits are summarized in Table \ref{beta}. The double $\beta$ model provides a lower reduced $\chi^{2}$ value than the single $\beta$ model. Therefore, we kept the double $\beta$ as the best model. We note that, in the analysis by \cite{a795}, a single $\beta$ model was found to provide a better fit to the {\it Chandra} data. This discrepancy could be due to the smaller field of view (FOV) of the {\it Chandra} observation, with a radius of 2.78$\arcmin$ (405 kpc), that may have prevented a reliable characterization of the outer $\beta$ model component.

\subsubsection{The southeast ICM extension: A possible cold front}

\label{excess2d}
\begin{figure*}
    \centering
    \begin{subfigure}{.45\textwidth}
        \includegraphics[height=7.5 cm]{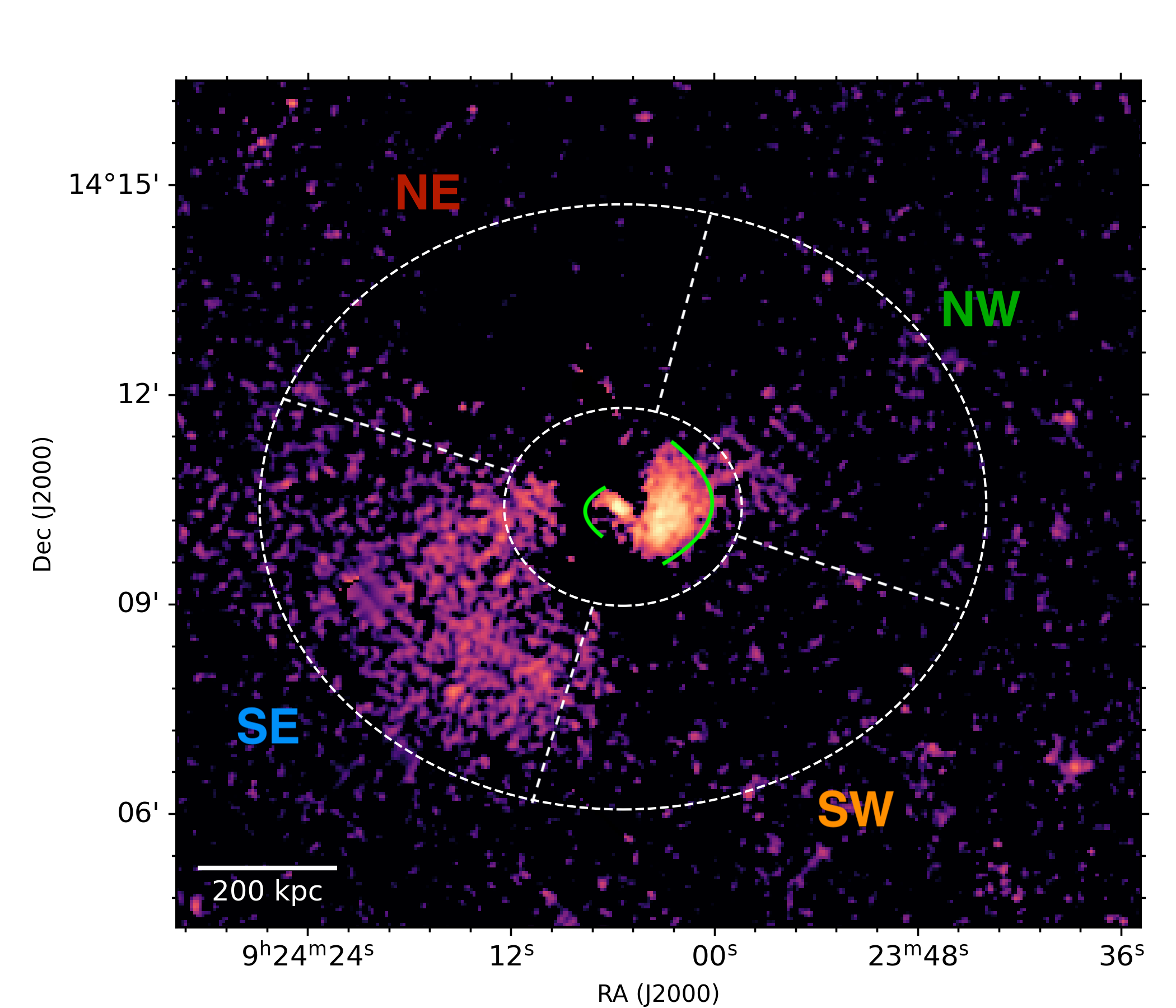}
    \end{subfigure}
    \hspace{0.3cm} %
    \begin{subfigure}{.5\textwidth}
        \includegraphics[height=7cm]{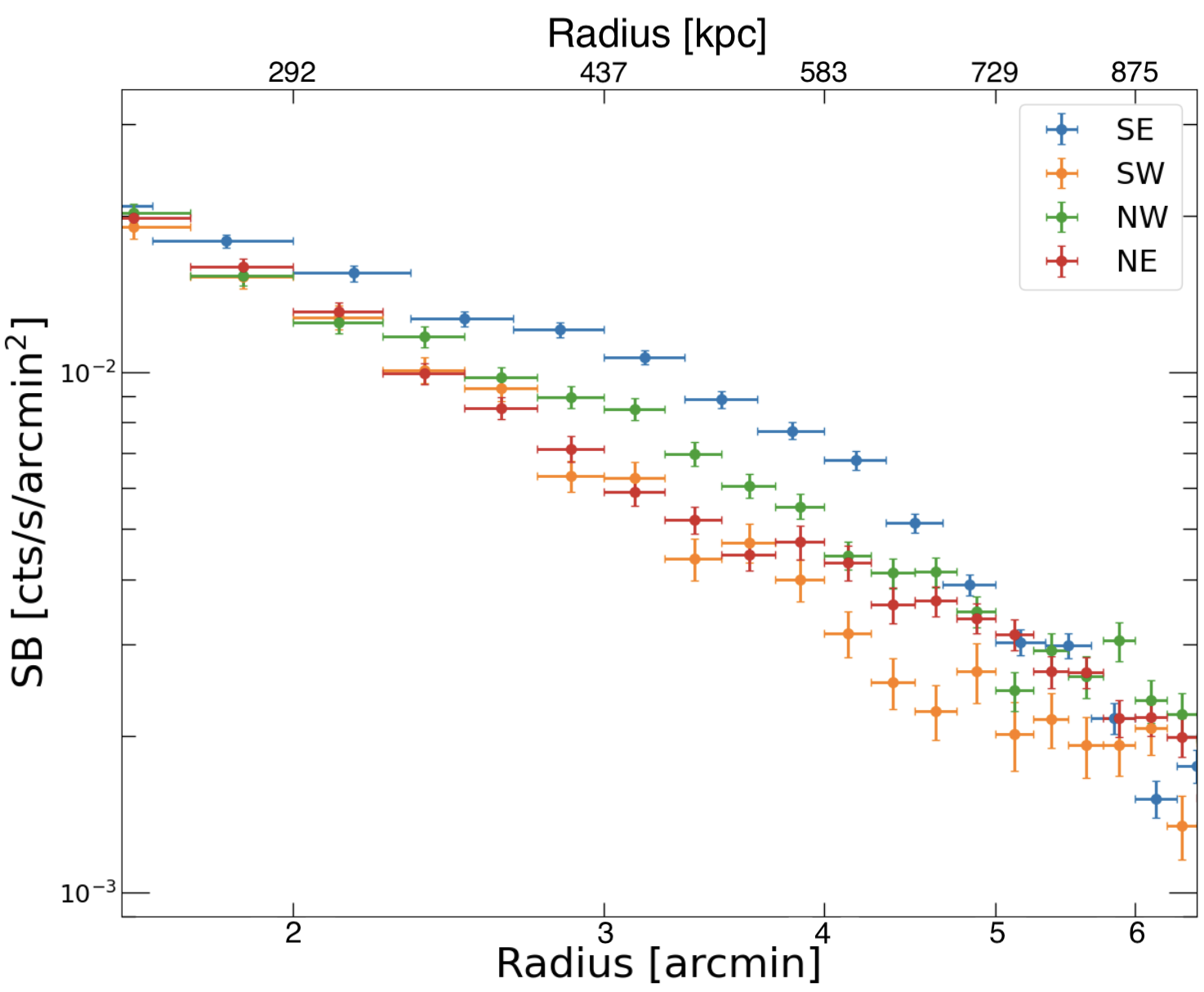}
    \end{subfigure}
    \caption{Morphological analysis of the large-scale surface brightness asymmetries in A795, see Sec. \ref{excess2d}. \textit{Left panel}: {\it XMM}-EPIC (0.5 -- 2 keV) residual image Gaussian-smoothed with a kernel radius of 9 pixels. The four regions mark the sectors used to extract the surface brightness profiles shown in the right panel. The inner and outer radii of the annulus correspond to 190 kpc and 1 Mpc from the X-ray peak, respectively. The two cold fronts detected by \cite{a795} are marked in green. \textit{Right panel}: {\it XMM}-EPIC (0.5 -- 2 keV) surface brightness profile of the four sectors. The blue points represent the SE sector which shows the excess.}
    \label{sectors}
\end{figure*}
In left panel of Fig. \ref{fluxim}, an asymmetric elongation of the X-ray emission toward the southeast direction can be observed. To highlight the asymmetry of the ICM, we obtained a residual image (left panel of Fig. \ref{sectors}) by subtracting the best-fit double $\beta$ model from the exposure-corrected {\it XMM} image of A795.
The residuals reveal a spiraling structure extending in the southeast direction, which appears to follow the previously detected sloshing motions. Two bright opposite regions are visible at the center of the cluster, which correspond to the cold fronts identified by \cite{a795} from {\it Chandra} data (green regions in left panel of Fig. \ref{sectors}).\\
With the aim of confirming the presence of the southeast asymmetric elongation of the ICM, we extracted the surface brightness profile in four elliptical sectors (each 90$^{\circ}$-wide, axis ratio b/a $=$ 0.83) centered on the X-ray peak. We focused on the radial range $r>290$ kpc (outside the inner two cold fronts found by \citealt{a795}) to $r\approx$ 1 Mpc. The resulting sectors are SE, NW, NE and SW, and are shown in left panel of Fig. \ref{sectors}. The right panel shows the SB profiles. The SE sector clearly shows an excess in surface brightness relative to the other directions, particularly from 2$\arcmin$ (292 kpc) to 6$\arcmin$ (875 kpc).
\begin{table*}[ht!]
\caption{Best-fit model of the X-ray surface brightness profile along the southest excess.}
    \centering
    \renewcommand{\arraystretch}{1.5}
    \begin{tabular}{c  c  c  c c  c  c }
        \hline
         & Model & \multicolumn{1}{c}{$\alpha_{1}$} & $\alpha_{2}$ & $r_{j}$ (kpc) & jump & $\chi^{2}$ / \textit{D.O.F.} \\
        \hline
        SE asymmetric extenion & Broken power-law & 1.27 $\pm$ 0.06 & 3.62 $\pm$ 0.20 & 650 $\pm$ 20 & 1.00 $\pm$ 0.15 & 21.4 / 12 (1.78)\\
        \hline
    \end{tabular}
    \label{bkn}
    \tablefoot{This table reports the best-fit parameters of the broken power-law model fitted to the X-ray surface brightness profile of the southeast asymmetric excess, see Fig. \ref{sectors}.}
\end{table*}
Given the spiral structure of the ICM and the presence of two inner cold fronts within the innermost $\approx$ 200 kpc of the cluster, it is possible that the southeastern asymmetric surface brightness excess represents a third, outer cold front. To verify this possibility, we considered a 3D broken power-law density model, defined in \textit{pyproffit} as
\begin{equation}
    F(r) = \begin{cases}
    r^{-\alpha_{1}} & \text{if } r < r_{j} \\
    \frac{1}{\text{jump}} \cdot r^{-\alpha_{2}} & \text{if } r \geq r_{j}
    \end{cases}\\,
\end{equation}
where "jump" defines the density jump across the front, $\alpha_{1}$ and $\alpha_{2}$ are the slopes before and after the jump, and $r_{j}$ is the distance of the jump from the center. This density model is projected along the line of sight and fitted to the observed surface brightness profile. The best-fit parameters are presented in Table \ref{bkn}. The jump parameter is consistent with unity, suggesting that this surface brightness excess cannot be classified as a contact discontinuity. Varying the surface brightness profile extraction region did not produce significant changes in the results. It is possible that, given the large distance of the excess from the center of the cluster ($\approx$ 650 kpc), deeper {\it XMM} observations would be necessary to verify the presence of a mild surface brightness discontinuity. The most cluster-center distant cold fronts detected so far in galaxy clusters are summarized in \cite{coldfar}. At distances larger than 500 kpc only five discontinuities are known. This highlights the difficulty in detecting such features at considerable distances from the center. However, it is not certain that all large-scale surface brightness excesses associated with sloshing should end with a cold front. Simulations predict the presence of surface brightness excesses at large distance from the cluster center but do not necessarily predict the formation of cold fronts in these regions (see Sec. 4.2 and Sec. 4.5 in \citealt{ross}).

\subsubsection{Detection of a candidate galaxy group}

\label{group2D}

By inspecting the exposure-corrected image, upper left panel in Fig. \ref{fluxim}, a region of enhanced, diffuse X-ray emission (above the surrounding background level) can be seen at $\approx$ 7.36$\arcmin$ northwest of the center of A795, as marked by the cyan dotted region.
We searched for known celestial objects around the approximate position of the X-ray diffuse emission in the Simbad archive\footnote{https://cds.unistra.fr}, finding a possible association with the radio galaxy NVSS J092352+141657 (RA, Dec = 09:23:52.75, +14:16:58.61). This galaxy has a redshift of z = 0.141, which is similar to that of A795 (z = 0.1374). In right panel of Fig. \ref{fluxim} the radio counter-part of this source is shown using the JVLA 1.5 GHz contours, while in the lower panel, we show an overlay of the 1.5 GHz radio contours with the DESI survey optical image. This optical image reveals four galaxies (including the optical host of NVSS J092352+141657) within 1.5\arcmin ($\approx$ 200 kpc) from the center of the diffuse X-ray emission, with similar spectroscopic redshifts (from \citealt{Rines}). To confirm the presence of extended X-ray emission at this position, we extracted the surface brightness profile from the {\it XMM} data using concentric circular annuli with linearly spaced binning of 6$\arcsec$, with an outer radius of 1.5$\arcmin$ (218.7 kpc), centered on the radio galaxy NVSS J092352+141657\\
Subsequently, we fitted the profile with a single $\beta$ model (see right panel in Fig. \ref{Sb}) and the best-fit parameters are summarized in Table \ref{beta} (last row).
The low reduced-$\chi^{2}$ value suggests that the single $\beta$ model is a good fit. The obtained $\beta$ value, $\beta$ = 0.53 $\pm$ 0.17, is close to the expected range for galaxy clusters or groups ($\beta \approx 2/3$, e.g., \citealt{betavalue}), strongly suggesting that the detected X-ray emission is produced by thermal plasma{\footnote{In this context, we also note that the radial surface brightness profile of the X-ray emission is significantly broader than that of the {\it XMM} PSF, further supporting the extended nature of the emission.}, such as the ICM or intragroup medium (IGrM). This, along with the detection of an elliptical galaxy with extended radio lobes (see right panel of Fig. \ref{fluxim}) at the center of the X-ray extended emission, provides further evidence supporting the classification of this object as a galaxy group. There are no previous mentions in the literature of galaxy groups present in the analyzed region. To better characterize the properties of this system, we present in Sec. \ref{group_analysis} a spectral analysis of its X-ray emission.\\
Furthermore, given the location of this candidate galaxy group in the proximity of a sloshing galaxy cluster, A795, we further explore in Sec. \ref{sloshing} any possible signs of past dynamical interactions between the two systems.\\

\section{Spectral analysis}
\label{spectral}
In this section, we presents the radio and X-ray spectral analysis. We first focus on the spectral index measurements of the different radio  components (the radio active BCG, the extended radio source) at the center of A795 detected in JVLA and GMRT data. We then report the results of the X-ray spectral analysis of the ICM of A795, as well as of the X-ray emission from the candidate galaxy group $\approx$ 1 Mpc northwest of the cluster.

\subsection{Radio spectral index analysis}

We measured the spectral index of the different sources by combining flux density measurements from the JVLA at 1.5 GHz and the GMRT at 325 MHz.
The derived flux density values are shown in Table \ref{fluxes} and were measured from the images in Fig. \ref{combined}, obtained with \texttt{robust} = 0.5. The resulting spectra are shown in Fig. \ref{alphaspectra}, with the values of $\alpha$ displayed in the bottom left corner.
\begin{figure}
    \centering
    \includegraphics[width=0.5\textwidth, height=0.5\textwidth]{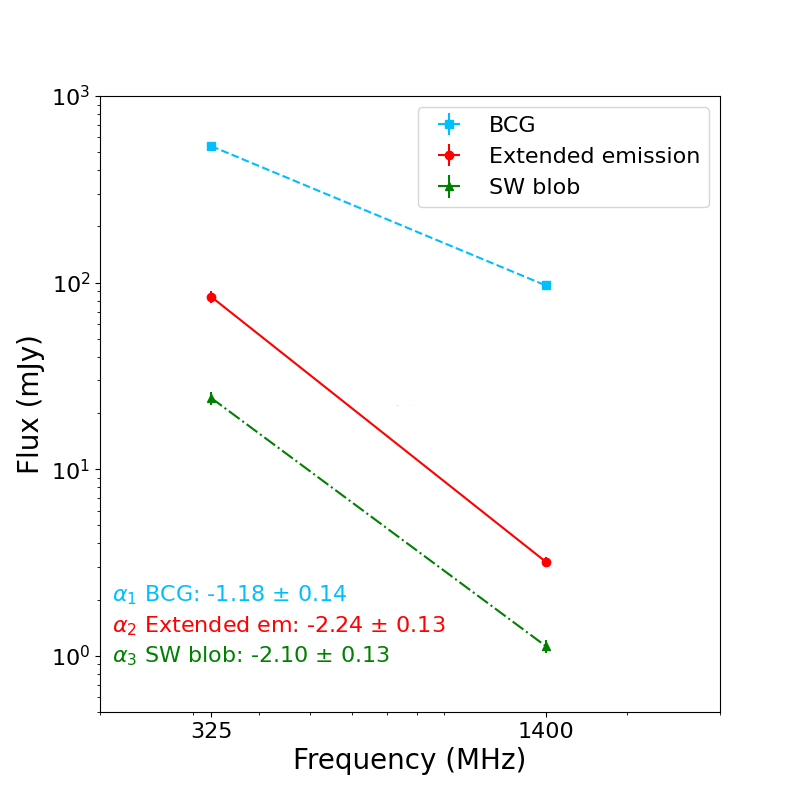}
    \caption{Radio spectra of the three radio sources detected at the center of A795, see Sec. \ref{radio}. In the bottom left corner, the spectral index $\alpha$ values are listed.}
    \label{alphaspectra}
\end{figure}
\begin{figure*}
    \centering
    \sidecaption
    \includegraphics[width=0.7\linewidth]{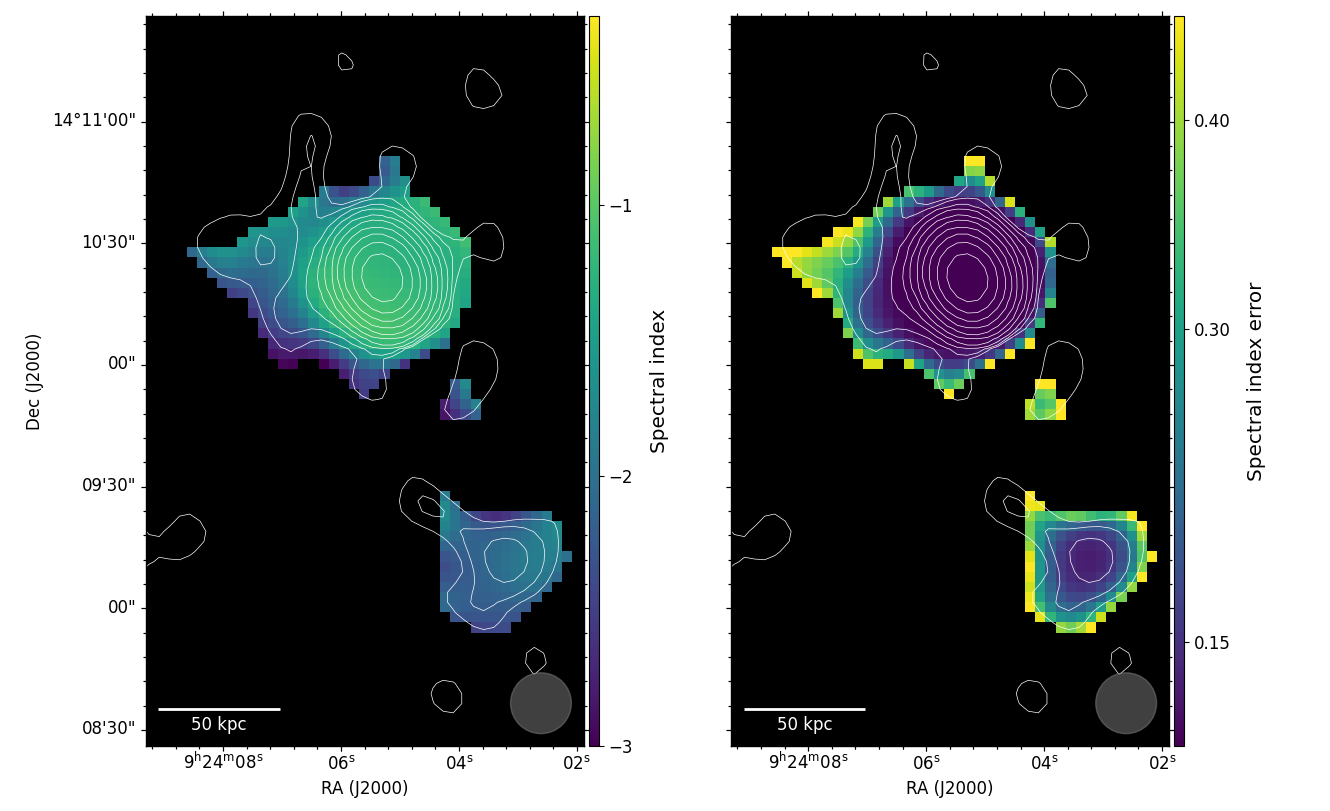}
    \caption{Spectral index map (left) and spectral index error map (right) measured at 325 MHz and 1.5 GHz of A795 zoomed on the central extended radio emission, with JVLA 1.5 GHz contours, defined in Fig. \ref{combined}.}
    \label{spix}
\end{figure*}
The BCG is an active radio galaxy and has a spectral index $\alpha_{BCG}$ = $-$1.18 $\pm$ 0.14, consistent with previous results at higher \citep[5 GHz and 8.4 GHz,][]{a795} and lower \citep[150 MHz and 325 MHz,][]{newa795} frequencies, who reported spectral indeces of $-$0.93 $\pm$ 0.09 and $-$1.08 $\pm$ 0.26, respectively.
The spectra we obtained for the extended emission ($\alpha_{Ext.}$ = $-$2.24 $\pm$ 0.13) and the SW blob ($\alpha_{SWb}$ = $-$2.10 $\pm$ 0.13), reported in Table \ref{spixpow}, are significantly steeper. For comparison, \cite{newa795} measured a spectral index $\alpha=-2.71\pm0.28$ for the extended emission\footnote{Given that the uncertainties are relatively large, and the fact that the regions we used to measure the flux densities are likely different from those of \cite{newa795}, these spectral indices can be considered broadly consistent.}. Such steep spectra are consistent with a scenario in which the extended emission originates from the reacceleration of populations of relativistic electrons. We further explore this aspect in Sec. \ref{phoenix}.\\
Additionally, we computed the radio power of the extended emission at 1.5 GHz as
\begin{equation}
    P_{1.5GHz} = 4 \pi D_{L}^{2} S_{1.5GHz} (1 + z)^{-(\alpha +1)}\\,
\end{equation}
where the factor (1+z)$^{-(\alpha + 1)}$ accounts for the K-correction.
The luminosity distance of A795 is $D_{L}$ = $671 \pm 49$ Mpc. Based on the 1.5 GHz flux density of the extended emission in Table \ref{fluxes}, the resulting radio power is $P_{1.5GHz}(Ext.)$ = $1.89 \pm 0.34 \cdot 10^{23}$ W/Hz.
We also computed the radio power of the SW blob using $\alpha_{SWb}$ = $-2.10 \pm 0.13$ and the value of $S_{1.5 GHz}$ from Table \ref{fluxes}, obtaining $P_{1.5GHz}(SW blob)$ = $0.63 \pm 0.12 \cdot 10^{23}$ W/Hz. The radio power values are summarized in Table \ref{spixpow}.
\begin{table}
\caption{Spectral index and radio power of the radio source components at the center of A795.}
    \centering
    \renewcommand{\arraystretch}{1.5}
    \begin{tabular}{c  c  c }
        \hline
         & \multicolumn{1}{c}{$\alpha$} & $P_{1.5 GHz}$ (W/Hz)\\
        \hline
        BCG & $-$1.18 $\pm$ 0.14  & 4.31 $\pm$ 0.68 $\cdot$ $10^{24}$\\
        Extended em. & $-$2.24 $\pm$ 1.3 & 1.89 $\pm$ 0.34 $\cdot$ $10^{23}$ \\
        SW blob & $-$2.10 $\pm$ 0.13 & 0.63 $\pm$ 0.12 $\cdot$ $10^{23}$ \\
        \hline
    \end{tabular}
    \label{spixpow}
\end{table}\\
In order to study the spectral index 2D distribution, we generated spectral index maps by combining the 1.5 GHz and 325 MHz radio images. To generate this, we used the CASA task $\texttt{immath}$, which calculates the spectral index values pixel by pixel. To perform this operation effectively, the two images must be perfectly aligned and have matching restoring beam sizes and uvrange. Different uvranges result in an unequal sampling of angular scales at the two frequencies involved. This produces a bias in the computation of $\alpha$. For this reason, we combined radio maps obtained by forcing a common uvrange of 0.2 -- 22 k$\lambda$, with a circular restoring beam of radius 15$\arcsec$. Figure \ref{spixmap} in the Appendix shows a wide-FOV spectral index map of A795, while Fig. \ref{spix} shows a zoom-in of the central region (as well as the spectral index error map). It can be seen how the spectral index steepens moving from the center of the emission, corresponding to the position of the BCG, toward the external regions. The SW blob shows a relatively uniform $\alpha$ distribution with steep values $\alpha \leq$ $-$2.

\subsection{X-ray spectral analysis}
\label{spec2dana}
\begin{table}
\caption{Best-fit parameters of the sky X-ray background from {\it XMM-Newton}.}
    \centering
    \renewcommand{\arraystretch}{1.5}
    \begin{threeparttable}
    \begin{tabular}{c  c  c }
        \hline
         & \multicolumn{1}{c}{\textit{norm}\tnote{a}} & kT (keV) \\
        \hline
        GH & 0.31 $\pm$ 0.02 & 0.28 $\pm$ 0.04 \\
        LHB & 3.09 $\pm$ 0.35 & --\\
        CXB & 1.14 $\pm$ 0.05 & --\\
        \hline
    \end{tabular}
    \begin{tablenotes}
            \item[a] In units of APEC XSPEC normalization per arcmin$^{2}$.
    \end{tablenotes}
    \end{threeparttable}
    \label{fit_bkg}
\end{table}
\begin{figure*}
    \centering
    \begin{subfigure}{.49\textwidth}
        \includegraphics[height=7.5 cm]{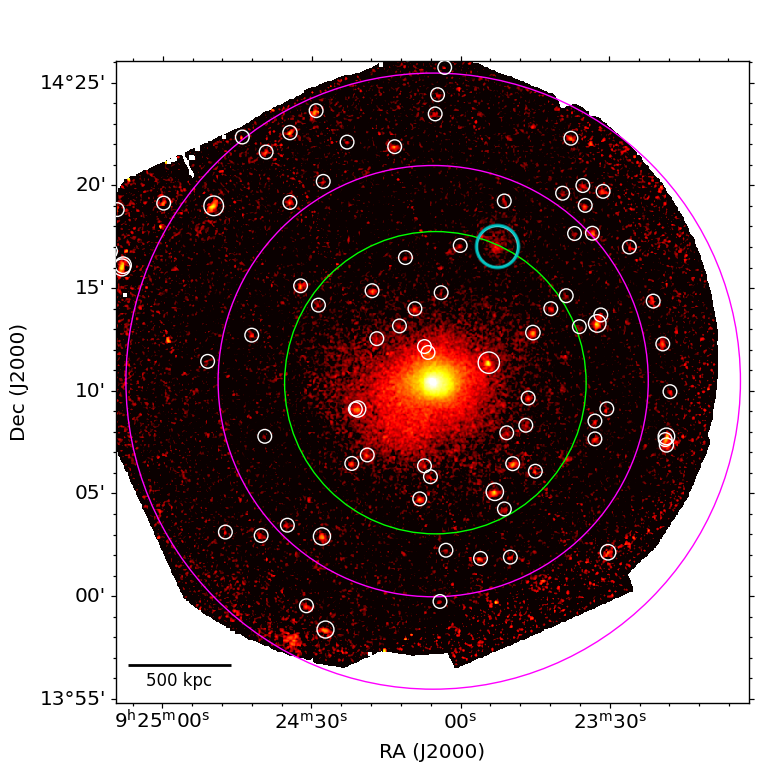}
    \end{subfigure}
    \hspace{0.01cm} %
    \begin{subfigure}{.5\textwidth}
        \includegraphics[height=7cm]{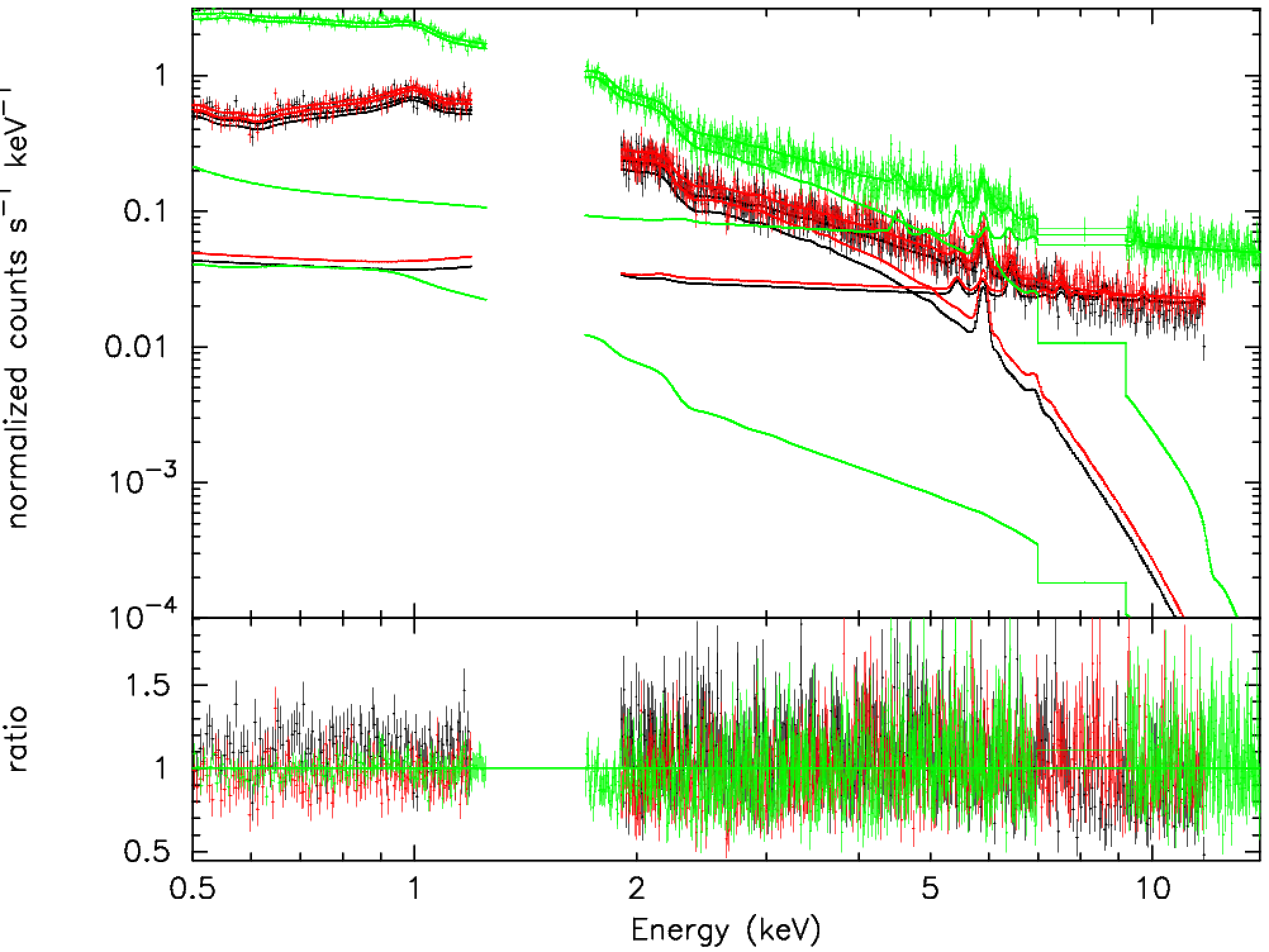}
    \end{subfigure}
    \caption{Spectral analysis of A795. \textit{Left panel}: {\it XMM}-EPIC (0.5 -- 2 keV) surface brightness image Gaussian-smoothed with a kernel radius of 3 pixels. The green region was used for the global spectral analysis of the cluster A795, while the magenta one for the background. The white circles mark the position of the excluded sources and the cyan region signs the position of A795-B, see Sect. \ref{group_analysis}. \textit{Right panel}: {\it XMM}-EPIC spectrum of A795 within the green region of the left panel, fitted with a thermal component and the background model. The red and black lines represent the models for MOS1 and MOS2 while the green one for pn. The lower panel shows the residuals from the best-fit curve. The best-fit parameters for the cluster thermal component are summarized in Table \ref{fit_combined}.}
    \label{spectrum}
\end{figure*}
To comprehensively characterize the thermodynamic properties of the galaxy cluster A795, we conducted an X-ray spectral analysis of the ICM. We adopted the approach of creating a background model rather than subtracting the background from the data, following, for example, \cite{lec} and \cite{snowden}. This differs from typical X-ray data analyses, where the background is subtracted. In the latter case, data taken from sky regions without sources called blank sky, are adapted and subtracted from the target observation. The issue with the subtraction method is the uncertainty regarding the background values that are subtracted, as they are derived from different sky regions and taken at different times. The method of creating a background model allows for greater control over its parameters, which are then measured on the observation of the source itself.
To fully characterize the background model, it is necessary to take into account all possible sources of detection. In the band 0.7 -- 10 keV the background spectral components can be categorized into two groups: the sky background and the particle background.
The sky background includes three components: (i) the Cosmic X-ray Background (CXB), which represents the integrated emission from all extragalactic sources in the X-ray band; (ii) the Local Hot Bubble (LHB), a region of ionized X-ray emitting gas surrounding the Sun; and (iii) the Galactic Halo (GH) component.
The particle background, also referred to as the non X-ray background (NXB), comprises high-energy cosmic ray particles of galactic origin (typically around GeV energies), and soft protons.\\
To study the global thermodynamic properties of A795, we employed a circular region centered on the X-ray peak, with a radius of R$_{500}$\footnote{R$_{500}$ is the cluster-centric distance within which the mean density is 500 times the critical density of the Universe at the cluster redshift.} = 7.4$\arcmin$ (1.02 Mpc). $R_{500}$ was estimated from the total mass $M_{500}$ value found in the Planck catalog, using Eq. 9 of \cite{r500}. For the background, we selected an annulus centered on the X-ray peak, with an inner radius corresponding to $R_{200}$ = 9.6$\arcmin$ (1.33 Mpc\footnote{Assuming an NFW profile, $R_{200}$ = 1.3 $R_{500}$.}), beyond which the cluster's emission is minimal, and an outer radius delineated by the limit of the {\it XMM} field of view, $\approx$ 15$\arcmin$ (2.19 Mpc). The spectral extraction regions are shown in left panel of Fig. \ref{spectrum}.\\
The extracted spectra were binned so that every bin contained 20 counts. We first modeled the background region. This enables obtaining the best-fit parameters for the various components of the background, which are then used in calculating the source spectrum. For a detailed description of the background model components refer to works such as \cite{pip3} and \cite{ross2024}.\\
For every thermal model and photoelectric absorption model employed in this work, we used the table of abundances of \cite{asplund}. The best-fit parameters for the sky background model are summarized in Table \ref{fit_bkg}. The model used for the cluster A795, which contains the model for the sky background, is \texttt{(apec$_{1}$+phabs*(apec$_{2}$+powerlaw+apec$_{3}$))} in XSPEC. The \texttt{apec$_{3}$} accounts for the cluster absorbed thermal emission and it has abundance, temperature and normalization let free to vary. The best-fit parameters for A795's thermal emission are shown in Table \ref{fit_combined}. Right panel in Fig. \ref{spectrum} shows the spectrum of A795 fitted with its model. The red and black data points are measured by MOS1 and MOS2, and the green ones by pn. The energy range $\approx$ 1.2 -- 2 keV is excluded for the three instruments due to the presence of strong fluorescence emission lines. For the pn, we also removed the range $\approx$ 6.8 -- 9 keV.\\
The abundance Z = 0.37 $\pm$ 0.02 $Z_{\odot}$ and the temperature kT = 3.87 $\pm$ 0.12 keV (4.49 $\pm$ 0.14 $\cdot$ $10^{7}$ K) are consistent with values typically found in galaxy clusters \citep{boh}. The global temperature is lower with respect to that reported in \cite{a795}, kT = 4.63 $\pm$ 0.12 keV, measured from {\it Chandra} data. It is important to mention that there exists a systematic difference in temperature measurement between analyses using {\it Chandra}/ACIS and {\it XMM}-EPIC data (full band), that is, EPIC data yielding 16$\%$ lower global temperatures at 5 keV \citep{Schellenberger_2015}. The difference observed here is consistent with this 16$\%$ offset, further supporting the systematic nature of the discrepancy. In addition to this, other factors may contribute to this temperature difference, including: the spectrum extraction region of {\it Chandra} data had a radius of 2.78$\arcmin$ (405.8 kpc), while the region used for the {\it XMM} analysis has a radius of 7.36$\arcmin$ (1.02 Mpc); the model used for the absorption  employed by \cite{a795} was \texttt{tbabs} in XSPEC, while in this work we used \texttt{phabs}; the column density value used for the absorption component is different since they used only the atomic hydrogen, while in our analysis also the molecular hydrogen was taken into account, from \cite{bourdin}. The column density parameter used in this analysis was fixed at the value N$_{H}$ = 4.35 $\cdot$ 10$^{20}$ cm$^{-2}$.

\subsection{Spectral analysis of the candidate galaxy group}
\label{group_analysis}

The X-ray spectral analysis of the candidate group named A795-B required a modification of the spectral fitting model, since the emission from the outer regions of the cluster A795 constitutes an additional background component in the group region. The steps taken are as follows:

\begin{itemize}
    \item Background region: We used the same annular region employed in the A795 background analysis for the A795-B background extraction (see the left panel of Fig. \ref{spectrum}). The best-fit parameters for the background components are reported in Table \ref{fit_bkg}.
    \item Cluster Emission: To account for the A795 emission at the candidate group's location, we added a thermal component to the source model. We derived the parameters for this thermal component by fitting the spectrum of an annular region centered on the X-ray peak of A795. The inner radius of the annulus was 5.85$\arcmin$ (852 kpc), while the outer radius was 8.33$\arcmin$ (1.21 Mpc), encompassing A795-B (see left panel of Fig. \ref{grp_spectrum}). During the extraction, the emission from A795-B was excluded. The best-fit parameters for this annular region were $kT = 2.53 \pm 0.12 \, \mathrm{keV}$, $Z = 0.17 \pm 0.07 \, Z_{\odot}$, and normalization $\textit{norm} = 1.03 \pm 0.21$ in units of APEC XSPEC normalization per arcmin$^{2}$.
    \item Group Emission: We extracted the spectrum of A795-B from a circular region centered on the associated radio galaxy, with a radius of 1.5$\arcmin$ (218.7 kpc; see left panel of Fig. \ref{grp_spectrum}). This radio galaxy lies 7.36$\arcmin$ (1.02 Mpc) from the cluster center, consistent with the $R_{500}$ of A795.
    \item Spectral Fitting: We modeled the A795-B spectrum in XSPEC as:
        \texttt{(apec}$_{1}$ + \texttt{phabs} * (\texttt{apec}$_{2}$ + \texttt{powerlaw} + \texttt{apec}$_{3}$+ \texttt{apec}$_{4}$)),
    where \texttt{apec$_{3}$} accounts for A795's thermal emission at the group's location, and \texttt{apec$_{4}$} represents the thermal emission of the group itself.
\end{itemize}
\begin{figure*}[ht!]
    \centering
    \begin{subfigure}{.45\textwidth}
        \includegraphics[height=7.5 cm]{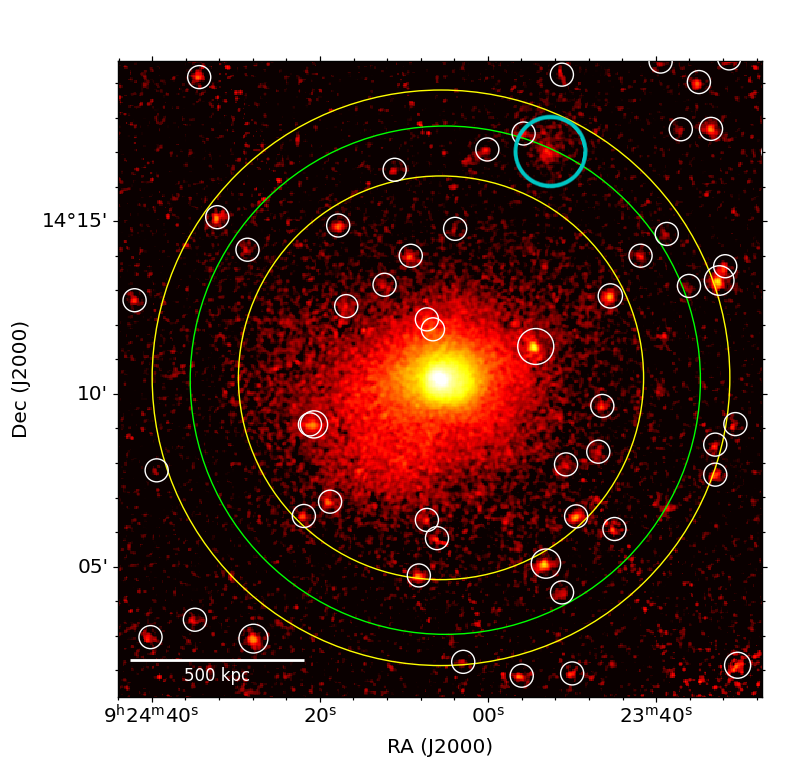}
    \end{subfigure}
    \hspace{0.1cm} %
    \begin{subfigure}{.5\textwidth}
        \includegraphics[height=7cm]{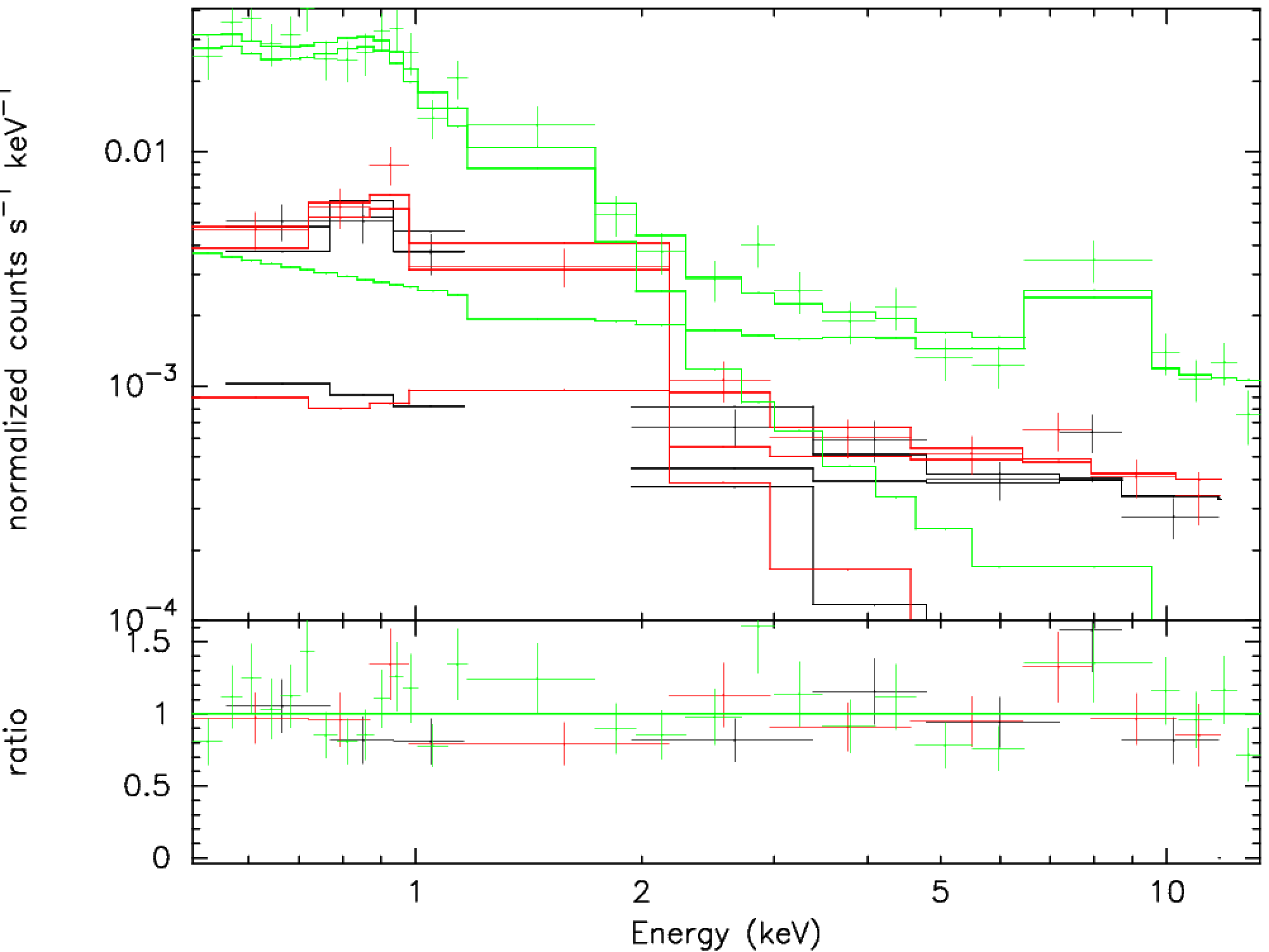}
    \end{subfigure}
    \caption{Spectral analysis of A795-B, see Sec. \ref{group_analysis}. \textit{Left panel}: {\it XMM}-EPIC (0.5 -- 2 keV) surface brightness image Gaussian-smoothed with a kernel radius of 3 pixels. The yellow annular region was used for the spectral extraction of the cluster component, while the green circle marks the position of $R_{500}$ = 1.02 Mpc of A795. The cyan region was used for the spectral extraction of A795-B. The white circles represent the point sources excluded. \textit{Right panel}: {\it XMM}-EPIC spectrum of A795-B fitted with a thermal component and the background model accounting for the cluster emission. The red and black lines represent the models for MOS1 and MOS2 while the green one for pn. The lower panel shows the residuals from the best-fit curve. The best-fit parameters for A795-B's thermal component are summarized in Table \ref{fit_combined}.}
    \label{grp_spectrum}
\end{figure*}
The spectrum is shown in the right panel of Fig. \ref{grp_spectrum}, and A795-B's best-fit parameters are summarized in Table \ref{fit_combined}.
The measured temperature of the system, kT = 1.02 $\pm$ 0.08 keV (1.18 $\pm$ 0.08 $\cdot$ $10^{7}$ K), agrees with expectations for galaxy groups \citep{bower}. This, along with the surface brightness profile compatible with a $\beta$ model, provides further confirmation that A795-B is a galaxy group.\\
\begin{table}
\caption{Best-fit parameters for the global X-ray spectral analysis of A795 and A795-B.}
    \centering
    \renewcommand{\arraystretch}{1.5}
    \setlength{\tabcolsep}{4pt}
    \begin{threeparttable}
    \begin{tabular}{c c c c c}
        \hline
        & \textit{norm}\tnote{a} & kT (keV) & Abund. ($Z_{\odot}$) & N$_{H}$ (cm$^{-2}$) \\
        \hline
        A795   & 6.42 $\pm$ 0.05 & 3.87 $\pm$ 0.12 & 0.37 $\pm$ 0.02 & 4.35 $\cdot$ 10$^{20}$\\
        A795-B & 1.71 $\pm$ 0.48 & 1.02 $\pm$ 0.08 & 0.13 $\pm$ 0.06 & 4.35 $\cdot$ 10$^{20}$\\
        \hline
    \end{tabular}
    \begin{tablenotes}
        \item[a] In units of APEC XSPEC normalization per arcmin$^{2}$.
    \end{tablenotes}
    \end{threeparttable}
    \tablefoot{The column density value was fixed during the analysis. Profiles are shown in the right panels of Fig. \ref{spectrum} and Fig. \ref{grp_spectrum}, respectively.}
    \label{fit_combined}
\end{table}
The \textit{norm} parameter is defined as
\begin{equation}
\textit{norm} = \frac{10^{-14}}{4\pi [D_A(1+z)]^2} \int n_e n_p dV\\,
\label{norm}
\end{equation}
where D$_{A}$ is the angular distance from the source, n$_{e}$ and n$_{p}$ are the electron and proton number densities, respectively, and V is the volume of the emitting region. The \textit{norm} parameter enables the calculation of the IGrM electron number density, which can be used to estimate its total mass. However, it is important to note that the \textit{norm} value, in this case, is derived from a spectrum that is not deprojected, meaning that the electron density may be overestimated due to projection effects.
By reverting Eq. \ref{norm} and considering n$_{e}$ = 1.2 n$_{p}$ (e.g., \citealt{Gitti}), it is possible to calculate the average electron number density n$_{e}$ in the group volume as
\begin{equation}
n_e = \sqrt{ \frac{10^{14} \left( 4\pi \cdot \text{norm} \cdot [D_A (1+z)]^2 \right)}{0.83 V}}\\,
\label{density}
\end{equation}
where V is the volume of the region used for the spectrum extraction of the group.
Assuming a radius of 1.5$\arcmin$ (218.7 kpc) from the morphological analysis, z = 0.141 and D$_{A}$ = 516 Mpc we get n$_{e}$ = 2.7 $\pm$ 0.6 $\cdot$ 10$^{-4}$ cm$^{-3}$, which is consistent with typical IGrM mass estimates for galaxy groups \citep{sun}.
The total IGrM density can then be expressed as $\rho_{IGrM}$ = 1.92 $\mu$ $\cdot$ n$_{e}$ $\cdot$ m$_{p}$, where  $\mu$ = 0.6 is the mean molecular weight and m$_{p}$ = 1.67 $\cdot$ 10$^{-24}$ g is the proton mass. Finally, a simple estimate of the IGrM mass is obtained from
M$_{IGrM}$ = $\rho_{IGrM}$ $\cdot$ V, leading to M$_{IGrM}$ (r $\leq$ 218 kpc) = 3 $\pm$ 1 $\cdot$ 10$^{11}$ M$_{\odot}$, which is a result typically expected for galaxy groups \citep{sun1}.\\
The total mass of galaxy clusters or groups can be estimated using the hydrostatic equilibrium assumption. The latter is based on the balance between gravity and pressure:
\begin{equation}
    \nabla P_{\text{gas}} = -\rho_{\text{gas}} \nabla \phi\\,
\end{equation}
where $P_{\text{gas}}$ is the gas pressure, $\rho_{\text{gas}}$ the gas density, and $\phi$ the gravitational potential. Under the assumption of spherically symmetric gas distribution and considering that  $P_{\text{gas}} = \rho_{\text{gas}} k T / \mu m_{p}$, it is possible to derive the expression for the mass profile M(r) inside the radius r as
\begin{equation}
        M_{tot}(<r) = -\frac{r k T}{\mu m_{p}} \left( \frac{d\ln\rho_{\text{gas}}(r)}{d\ln(r)} + \frac{d\ln T(r)}{d\ln(r)} \right)\\,
\label{hydrostatic}
\end{equation}
where $\mu$ is the mean molecular weight, $m_{p}$ is the proton mass, and k is the Boltzmann constant.
Since the single $\beta$ model was used to fit the X-ray surface brightness of A795-B, the total mass can be recovered analytically and expressed by the formula
\begin{equation}
M_{tot}(< r) = \frac{k \, r^2}{G \, \mu \, m_p} \left[ \frac{3 \beta r T}{r^2 + r_c^2} - \frac{dT}{dr} \right]\\.
\end{equation}
Substituting the single $\beta$ model best-fit parameters from Table \ref{beta} (last row), the gravitational constant G = 6.67 $\cdot$ 10$^{-8}$ cm$^{3}$/g/s$^{2}$, the Boltzmann constant k = 1.38 $\cdot$ 10$^{-16}$ erg/K, r = 1.5$\arcmin$ (218.7 kpc), T = 4.49 $\cdot$ 10$^{7}$ K and assuming a constant temperature profile ($\frac{dT}{dr} = 0$) we obtained the gravitational mass of A795-B M$_{tot}$ (r $\leq$ 218 kpc) = 7 $\pm$ 3 $\cdot$ 10$^{12}$ M$_{\odot}$. This result is consistent with predictions for galaxy groups \citep{Tully}. As a note of caution, we point out that the above M$_{IGrM}$ and M$_{tot}$ were derived within a fixed radius of 218~kpc (the extent of the group in X-rays, see Fig. \ref{Sb}, right), rather than within a scaled radius (e.g., R$_{500}$), and thus the comparison with reference values for groups is only approximate.

\section{Discussion}
\label{discussion}
\begin{figure*}[ht!]
    \centering
    \includegraphics[width=\linewidth]{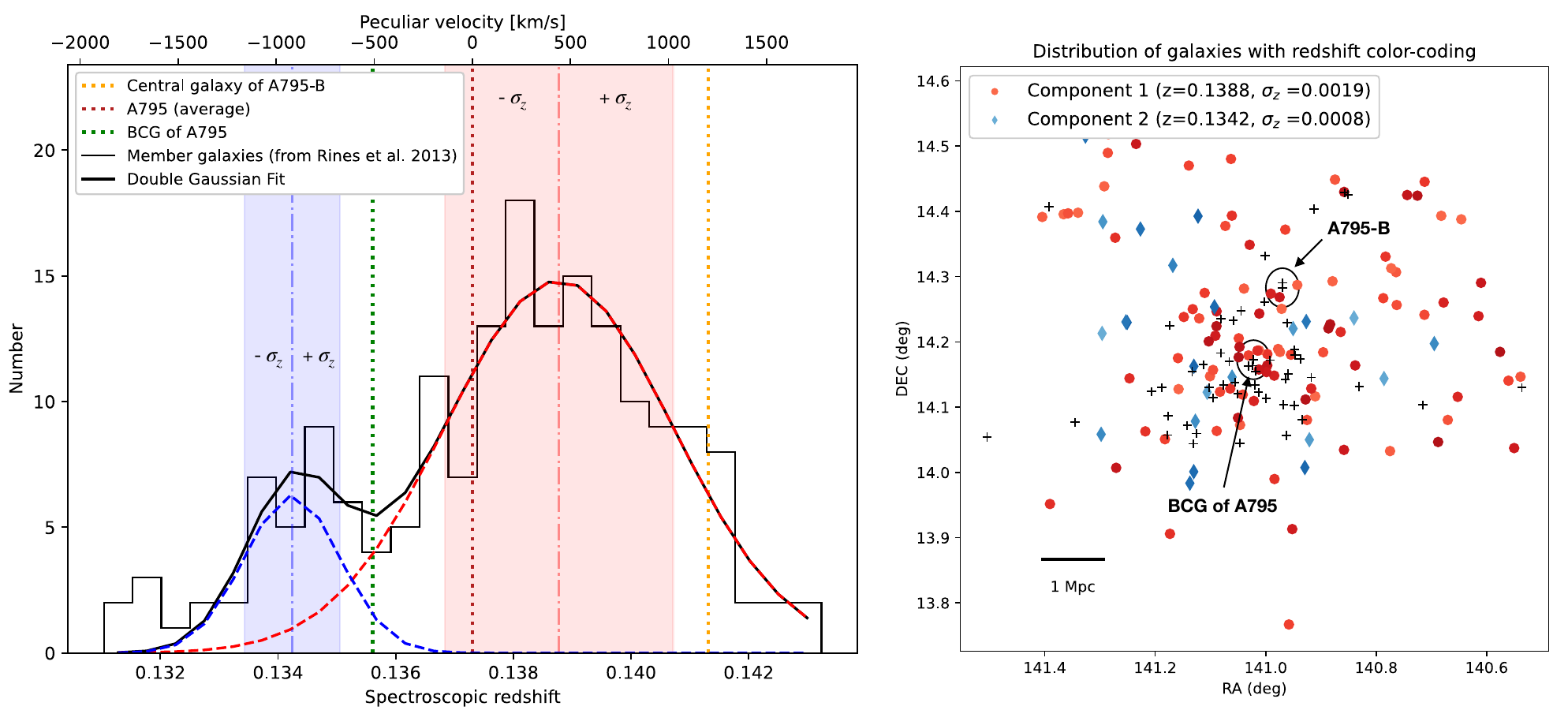}
    \caption{Velocity distribution of member galaxies in A795. {\it Left panel}: histogram of the spectroscopic redshift (bottom x-axis) for 179 member galaxies of A795, measured and reported in \citet{Rines}. The top x-axis shows the peculiar velocity relative to the average redshift of A795, that is z = 0.1374. Vertical dotted lines mark the central galaxy of A795-B (yellow), the average of A795 (red), and the BCG of A795 (green). The black solid line is the best-fit with a double Gaussian of the histogram, with the two components shown as blue and red dashed lines. The vertical dashed-dotted lines show the central velocity of each gaussian, while the shaded colored area represents the 1$\sigma$ width of each Gaussian. {\it Right panel}: spatial distribution in RA and DEC of the member galaxies of A795 from \citet{Rines}, color coded based on their peculiar velocity. Galaxies represented by red (blue) circles have peculiar velocity within $\pm1\sigma$ from the corresponding average of z = 0.1388 (z = 0.1342) shown in the left panel. Black + signs represent galaxies whose peculiar velocity falls outside the $\pm1\sigma$ range of both Gaussians in the left panel. The spatial position of the BCG of A795 and of the central galaxy of A795-B is highlighted, and both are plotted as with a black + sign.}
    \label{optical_velocity}
\end{figure*}
The radio analysis of A795 with the new JVLA data at 1.5 GHz and the archival GMRT image at 325 MHz revealed the presence of an extended synchrotron source with elongated morphology, surrounding the central BCG. We discuss the classification and origin of this emission in Sec. \ref{phoenix}.\\
The new {\it XMM}-EPIC (0.5 -- 2 keV) surface brightness analysis revealed two extended features. The first is an arc-shaped surface brightness excess in the southeast region that appears to follow the sloshing spiral previously detected by \cite{a795} (see the next section for a comprehensive analysis). The second is an extended source located northwest of the cluster center. The JVLA image at 1.5 GHz reveals an extended radio galaxy coincident with this feature. The surface brightness profile well fitted by a single $\beta$ model and the presence of thermal IGrM at temperature kT = 1.02 $\pm$ 0.08 keV (1.18 $\pm$ 0.08 $\cdot$ $10^{7}$ K), measured for the first time from the spectral analysis, corroborate the hypothesis that this object can be identified as a galaxy group, that we named A795-B. In the next section, we explore whether A795-B could be the perturber responsible for initiating the sloshing motion.

\subsection{Sloshing in A795}

\label{sloshing}

Evidence of sloshing at the center of A795 was previously reported using {\it Chandra} observations \citep{a795}, which revealed two cold fronts located at 59.6 $\pm$ 0.3 kpc and 178.2 $\pm$ 2.2 kpc from the cluster center (see also \citealt{newa795} for consistent results). With {\it XMM}-EPIC data we discovered a surface brightness arc-shaped excess reaching $\approx$ 650 kpc from the center, that seems to follow the sloshing spiral (see left panel in Fig. \ref{sectors}). However, this excess could not be classified as a density discontinuity since it did not show a significant edge in the SB profile. Such features have been qualitatively reproduced in simulations of galaxy clusters and groups, and can further constrain the geometry of sloshing (e.g., \citealt{roediger2011}, \citealt{roediger2012}).\\
A similar configuration was observed in the galaxy group NGC 5044, where an X-ray surface brightness excess at 110 kpc from the center, which could not be confirmed as a proper third cold front, was found following the spiral structure traced by two inner cold fronts \citep{gastald}. In the case of A795, the surface brightness excess is far more distant from the center ($\approx$ 650 kpc). As noted earlier, only five discontinuities have been identified at distances beyond 500 kpc from the center of their host cluster center \citep{coldfar}. This underscores the challenges associated with resolving these morphological substructures in the peripheral regions of galaxy clusters, primarily due to the rapid outward decline in surface brightness. Additionally, simulations indicate that large-scale surface brightness excesses, while common in sloshing systems, do not always coincide with the formation of cold fronts at such distances from the cluster center \citep{ross}. \\
The galaxy group we detected, A795-B, could be the candidate perturber for originating the sloshing motion. Simulations of gas sloshing in galaxy clusters, such as the one conducted by \cite{ascasibar}, show how as the perturber falls in the gravitational potential well, its outer gas is stripped and forms a comet-shaped tail. Observational confirmations of this scenario can be found in the galaxy cluster RXJ1347.5-1145 \citep{johnson}, one of the few objects for which the perturber galaxy group could be identified. This galaxy group exhibited a low gas content, likely indicating that a previous passage near the cluster center had depleted its gas. We therefore consider the possibility that A795-B is the perturber of A795. Using the redshift of the group (z = 0.141) and the average redshift of the cluster (z = 0.1374), we can measure the line of sight component of the peculiar velocity of A795-B with respect to A795, finding a reasonable value of v $\approx$ 1200 km/s (the velocity dispersion of A795 is $\approx770$~km/s, see \citealt{Rines}). The group has a gravitational mass M$_{tot}$ (r $\leq$ 218 kpc) = 7 $\pm$ 3 $\cdot$ 10$^{12}$ M$_{\odot}$ and presents IGrM with mass M$_{IGrM}$ (r $\leq$ 218 kpc) = 3 $\pm$ 1 $\cdot$ 10$^{11}$ M$_{\odot}$, values that are in line with the typical gas fraction \textit{f}$_{gas}$ for galaxy groups \citep{sun}. However, the gas fraction alone is insufficient to rule out a past interaction, as seen, for instance, in \cite{Ichinohe}. Specifically, their study of the galaxy cluster Abell 85 reveals that it is undergoing a double merger, with one of the two infalling substructures exhibiting an X-ray bright core. Moreover, we did not detect any sign of gas stripping forming a comet-shaped tail around the group. This could be attributed to the exposure time of the {\it XMM} dataset analyzed in this work, projection effects, or a combination of both.\\
\begin{figure*}[ht!]
    \centering
    \sidecaption
    \includegraphics[width=0.7\textwidth]{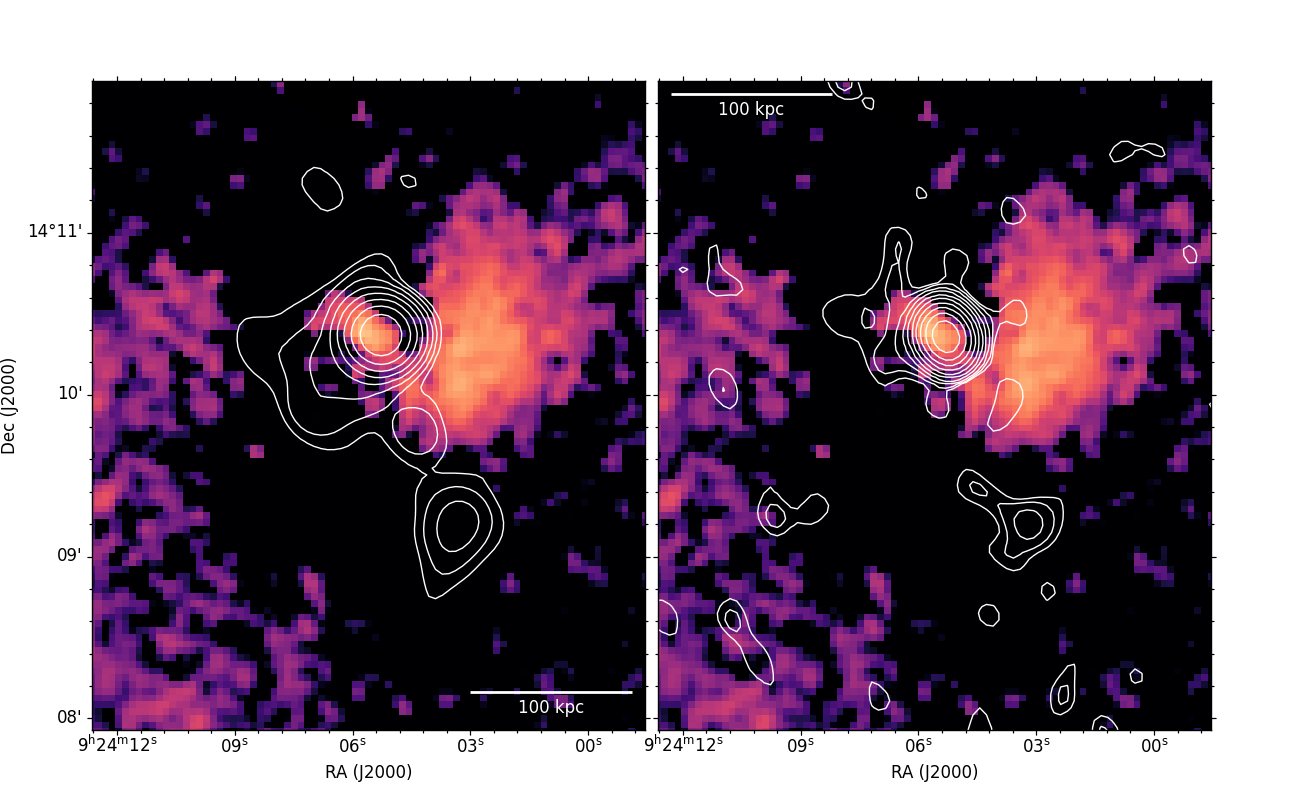}
    \caption{{\it XMM}-EPIC (0.5 -- 2 keV) residual image Gaussian-smoothed with a kernel radius of 9 pixels of A795's central region  with GMRT 325 MHz contours (left) and JVLA 1.5 GHz contours (right), defined in Fig. \ref{combined}.}
    \label{combined_residual}
\end{figure*}
For completeness, we comment on the velocity distribution of member galaxies in A795. Past studies have shown that the spatial and kinematical distribution of member galaxies can be useful to reveal the presence of substructures (e.g., \citealt{girardi,gastald,brienza2022,benavides}). The identification of member galaxies of A795 has been performed by \citet{Rines}, who presented spectroscopic redshifts for 179 member galaxies of the cluster. Starting from their published catalog of spectroscopic redshifts, we plot in Fig. \ref{optical_velocity} the redshift and velocity distribution of the member galaxies. The histogram of the spectroscopic redshifts (left panel) reveals several signs of dynamical disturbances. First, the member galaxies are distributed in a clearly bimodal distribution ranging between $0.132\leq z \leq 0.142$. A simple fit with a double Gaussian distribution yields a 99.4\%-significance improvement with respect to a single Gaussian. One component is found to be located at z = $0.1342\pm0.0003$ (blue lines in Fig. \ref{optical_velocity}, left), with $\sigma_{v} = 240\pm77$~km/s ($\delta_{z} = 0.0008\pm0.0003$), and the other is located at z = $0.1388\pm0.0002$ (red lines in Fig. \ref{optical_velocity}, left), with $\sigma_{v} = 582\pm58$~km/s ($\delta_{z} = 0.0019\pm0.0002$). Interestingly, the member galaxies belonging to the blueshifted component and those belonging to the redshifted component are spatially coincident (Fig. \ref{optical_velocity}, right). This is consistent with a possible line-of-sight minor merger occurring in A795. Second, the BCG of A795 is at $v_{rel} = -500$~km/s with respect to the average of A795. This configuration is typical of sloshing galaxy clusters whose potential well has been perturbed (e.g., \citealt{coziol}), setting the BCG, the galaxies, and the ICM in relative motion. Notably, the BCG of A795 and the central galaxy of A795-B lie beyond $\pm1\sigma_{v}$ of the two Gaussians shown in Fig. \ref{optical_velocity} (left). Although a complete determination of the number and spatial distribution of substructures in A795 is beyond the scope of this work, the simple plots in Fig. \ref{optical_velocity} already demonstrate the presence of velocity substructures in the cluster, as well as their likely alignment with the line of sight. Therefore, while A795-B is a candidate perturber, there are other structures within 10~Mpc of the center of A795 that might have set the sloshing oscillation of the ICM in motion.

\subsection{A candidate radio phoenix}

\label{phoenix}

In a recent study of A795, \cite{newa795} analyzed archival GMRT data at 150 MHz and 325 MHz and detected the extended radio emission with a complex shape at the center of the cluster. They interpreted this emission as a candidate mini-radio halo, even though they measured an ultra-steep spectral index of $-2.71 \pm 0.28$. This value differs from the typical range for these type of sources, which  is $-$1.4 < $\alpha$ < $-$1.1 \citep{Giovannini}. Moreover, they noted that the complex morphology poses challenges to the mini-radio halo interpretation, as these sources exhibit a rounder shape \citep{gittimini}.\\
In this work, we favor a different interpretation. The observed ultra-steep values of the spectral index  ($\alpha_{Ext.}$, $\alpha_{SWb}$ $\approx$ $-$2.2), the complex shape (see Fig. \ref{combined} and Fig. \ref{combined_natural} and the cospatiality with the BCG, which is a radio-loud AGN (P$_{1.5GHz}$ = 4.31 $\pm$ 0.68 $\cdot$ 10$^{24}$ W/Hz), align with the scenario of a revived AGN plasma source. These sources are characterized by steep radio spectral indeces, with values $\alpha$ $\leq$ $-$1.5 \citep{degasp1}, and complex morphologies predicted by simulations \citep{Enblin}. We also note that phoenices are expected to exhibit high-frequency spectral curvature, and not only a simple power-law spectrum \citep{adiabatic}. Specifically, this occurs because adiabatic compression revives fossil radio emission by shifting the original radio spectrum to higher frequencies. The original spectrum is already steep and curved due to the radiative losses the plasma experienced before the compression occurred. In the future, it would be desirable to obtain four or more flux density measurements to  probe any spectral curvature of the radio source in A795.\\
Among the key questions concerning radio phoenices, the most intriguing concerns the mechanism responsible for the reacceleration of the aged plasma emitted by the central AGN. Adiabatic compression has been proposed as a plausible explanation, attributed to the effects of shock passage. This mechanism effectively re-energizes electrons, leading to synchrotron emission characterized by a steep radio spectral index \citep{adiabatic}. However, a connection between the diffuse emission and shocks waves is missing,  as in the case of A795 no shock front was detected.\\
Another reacceleration scenario was proposed by \cite{botteon} for the case of Abell 2567. The detected diffuse radio emission was cospatial with the sloshing spiral at the center of the cluster (see their Fig. 5). Comparison with tailored numerical simulations (see also \citealt{zuhone1}) supported the idea that the relativistic AGN plasma may have been compressed, and re-energized by the sloshing motion.
The diffuse emission in A795 does not show, as in the case of Abell 2567, a perfect cospatiality with the enhancement in surface brightness caused by the sloshing (see Fig. \ref{combined_residual}), especially in the case of the SW blob, but this does not exclude a possible interaction (see the case of IC1860, \citealt{gastald}). Projection effects further complicate the picture, since they represent a possible bias for the system configuration.\\
If sloshing motion were responsible for the reacceleration, it is expected that tangential motions would stretched magnetic fields along the sloshing spiral. As a result, the presence of polarized emission is predicted  due to the alignment of magnetic field lines. Only a few polarization measurements have been performed so far of these sources \citep{slee}, finding low polarization levels. However, since phoenices are typically located near the cluster center, as in the case of A795, Faraday depolarization could significantly reduce the observed polarization fraction, especially at low frequencies. A polarization analysis of the diffuse emission detected in A795 at frequencies higher than 1.5 GHz could serve as a topic for future studies.

\section{Conclusions}
\label{conclusions}
In this work we presented the analysis of new JVLA 1.5 GHz observations, along with archival GMRT 325 MHz and XMM-Newton data, of the galaxy cluster A795.
The main results are summarized below.
\begin{itemize}
    \item The X-ray morphological analysis based on {\it XMM} data was conducted on scales larger than those covered with {\it Chandra} (\citealt{a795}, \citealt{newa795}), allowing us to study A795 out to  r$_{200}$ = 10.5$\arcmin$ (1.53 Mpc) (see Sec. \ref{spec2dana}). A double $\beta$ model was found to provide the best-fit to the surface brightness profile, with best-fit parameters of $\beta$ = 0.59 $\pm$ 0.03, r$_{c,1}$ = 59.8 $\pm$ 6.7 kpc and r$_{c,2}$ = 317.8 $\pm$ 34.5 kpc (see Table \ref{beta}, middle row). We determined the global properties of the thermal ICM in A795 through spectral analysis (see Sec. \ref{spec2dana}), measuring a temperature kT = 3.87 $\pm$ 0.12 keV and abundance Z = 0.37 $\pm$ 0.02 Z$_{\odot}$ (see Table \ref{fit_combined}).
    \item The X-ray surface brightness analysis revealed an arc-shaped, azimuthally asymmetric excess extending to $\approx$ 650 kpc from the center (see left panel in Fig. \ref{sectors}), which appears to follow the known sloshing spiral (see Sec. \ref{excess2d}). Deeper future {\it XMM} observations may determine whether this excess corresponds to a large-scale cold front.
    \item The {\it XMM} image revealed an extended source located northwest of the cluster center at a projected distance of $\approx$ 7.36$\arcmin$ (1.02 Mpc) from A795's center (see Fig. \ref{fluxim}), centered on the radio galaxy NVSS J092352+141657 (z = 0.141). Its X-ray surface brightness profile is well-fit by a $\beta$ model with best-fit parameters $\beta$ = 0.52 $\pm$ 0.17 and r$_{c}$ = 28.2 $\pm$ 13.1 kpc (see Table \ref{beta}). We identified this source as a candidate galaxy group. Extended radio emission has been found from the JVLA map at 1.5 GHz, associated with the  elliptical galaxy at the center of the group (see right panel in Fig. \ref{fluxim}). The X-ray spectrum supports the galaxy group interpretation, being well described by a thermal gas with temperature of kT = 1.02 $\pm$ 0.08 keV. We named it A795-B.
    \item We examined whether A795-B may have acted as the perturber that triggered the sloshing motion in A795 (see Sec. \ref{sloshing}). The peculiar velocity between the group and A795 (v $\approx$ 1200 km/s) is consistent with a high-speed core passage, but the gas fraction of the group and the lack of a clear cometary IGrM morphology may argue against a past passage close to A795's core. We conclude that A795-B is a plausible perturber, although further investigation is needed. The optical distribution of member galaxies of A795 shows evidence for sub-halos aligned along the line of sight, supporting the unrelaxed nature of A795.
    \item The new JVLA 1.5 GHz map confirmed the presence of extended emission surrounding the BCG of A795, with largest linear size $\approx$ 170 kpc, along with a sub-component extending in the southwest direction, named the SW blob, at a distance $\approx$ 200 kpc from the center (see right panel in Fig. \ref{combined}). This emission was also detected in GMRT archival data at 325 MHz, with an extension of $\approx$ 217 kpc (see left panel in Fig. \ref{combined}). We measured ultra-steep spectral indices of $\alpha_{Ext.}$ = $-$2.24 $\pm$ 0.13 for the extended emission (excluding the BCG) and $\alpha_{SWb}$ = $-$2.10 $\pm$ 0.13 for the SW blob (see Table \ref{spixpow}). These ultra-steep indeces, combined with the elongated complex morphology and cospatiality with the radio-loud AGN present in the BCG, suggest that the extended emission of A795 is a radio phoenix (see Sec. \ref{phoenix}). The reacceleration mechanism remains unclear, as the radio emission shows only a partial spatial match with the sloshing spiral in the ICM.
\end{itemize}
Future observations in polarization of the candidate radio phoenix may shed further light on this peculiar system and determine wether the magnetic field lines are aligned with the sloshing spiral -- providing evidence for sloshing-related reacceleration -- or not. In addition, higher-frequency radio observations could test wether the spectrum exhibits curvature above GHz frequencies, which would support an adiabatic compression scenario. Moreover, deeper {\it XMM} observations could enable the study of large-scale sloshing motion as well as any potential cometary morphology of the IGrM in A795-B.

\begin{acknowledgements}
We thank the referee for their clear revision of this manuscript. FU and MG acknowledge support from the research project PRIN 2022 ``AGN-sCAN: zooming-in on the AGN-galaxy connection since the cosmic noon", contract 2022JZJBHM\_002 -- CUP J53D23001610006. The National Radio Astronomy Observatory is a facility of the National Science Foundation operated under cooperative agreement by Associated Universities, Inc. We thank the staff of the GMRT that made these observations possible. GMRT is run by the National Centre for Radio Astrophysics of the Tata Institute of Fundamental Research.
\end{acknowledgements}

\bibliographystyle{aa}

\begin{appendix}
\onecolumn
\section{Radio images}
In this section we present three different views of the radio emission in A795. The first, Fig. \ref{combined_natural}, shows the central diffuse emission at 325 MHz and 1.5 GHz using natural weighting. In Fig. \ref{fr0} are presented the images at 325 MHz and 1.5 GHz that we used to measured the BCG flux, by excluding the contribution of extended sources. Figure \ref{spixmap} shows the wide field spectral index map of A795.
\begin{figure*}[h]
    \centering
    \sidecaption
    \includegraphics[width=0.7\linewidth]{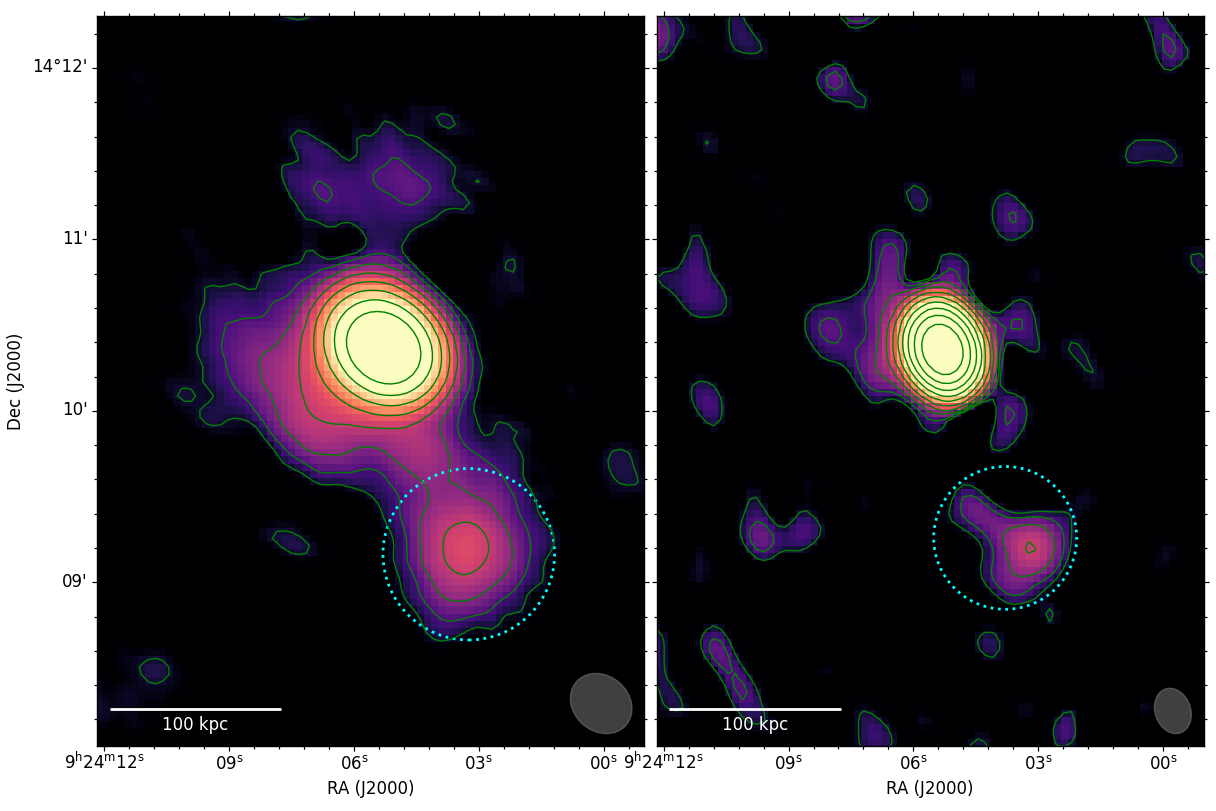}
    \caption{Radio images of the central radio source in A795 obtained with \texttt{briggs} = 2. \textit{Left panel}: GMRT 325 MHz image; the restoring beam of $23\arcsec \times 19\arcsec$ is shown in gray. RMS is 450 $\mu$Jy/beam. \textit{Right panel}: JVLA 1.5 GHz image; the restoring beam of $16.2\arcsec \times 12.5\arcsec$ is shown in gray. RMS is 27 $\mu$Jy/beam. In both panels, the contours are drawn at 3, 6, 12, 24, 48 $\times$ RMS and the dotted circle marks the SW blob.}
    \label{combined_natural}
\end{figure*}
\begin{figure}[h]
    \centering
    \sidecaption
    \includegraphics[width=0.7\linewidth]{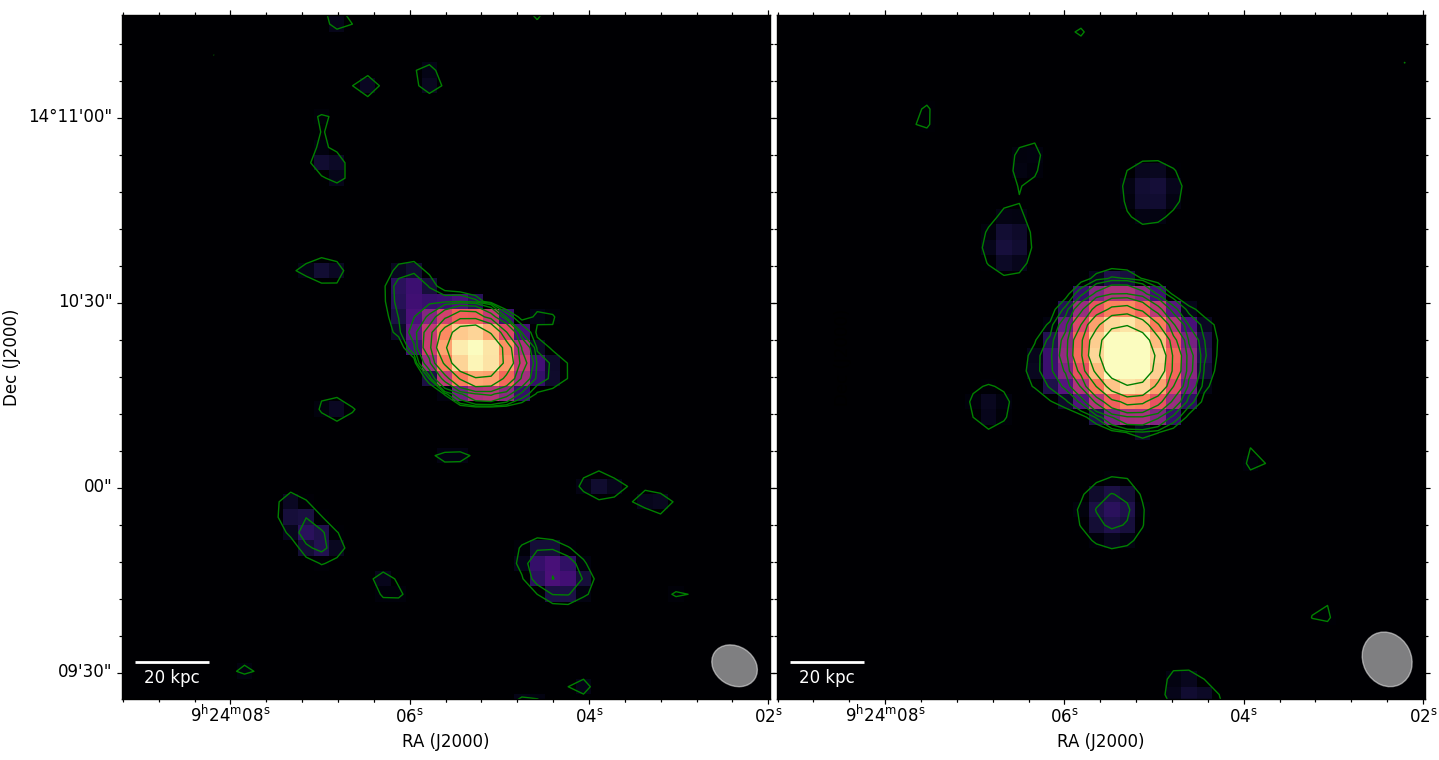}
    \caption{Radio images of the central source in A795 with a cut at baselines < 7k$\lambda$ made to select only the BCG emission. \textit{Left panel}: GMRT 325 MHz image obtained with \texttt{briggs} = 0.5 of the central region of A795. The restoring beam of $7.8\arcsec \times 6.3\arcsec$ is shown in gray. RMS is 480 $\mu$Jy/beam. \textit{Right panel}: JVLA 1.5 GHz image obtained with \texttt{briggs} = 0.5. The restoring beam of $9.1\arcsec \times 7.8\arcsec$ is shown in gray. RMS is 54 $\mu$Jy/beam. In both panels, the contours are at levels 3, 6, 12, 24, 48 RMS.}
    \label{fr0}
\end{figure}
\begin{figure}[h]
    \centering
    \sidecaption
    \includegraphics[width=0.7\linewidth]{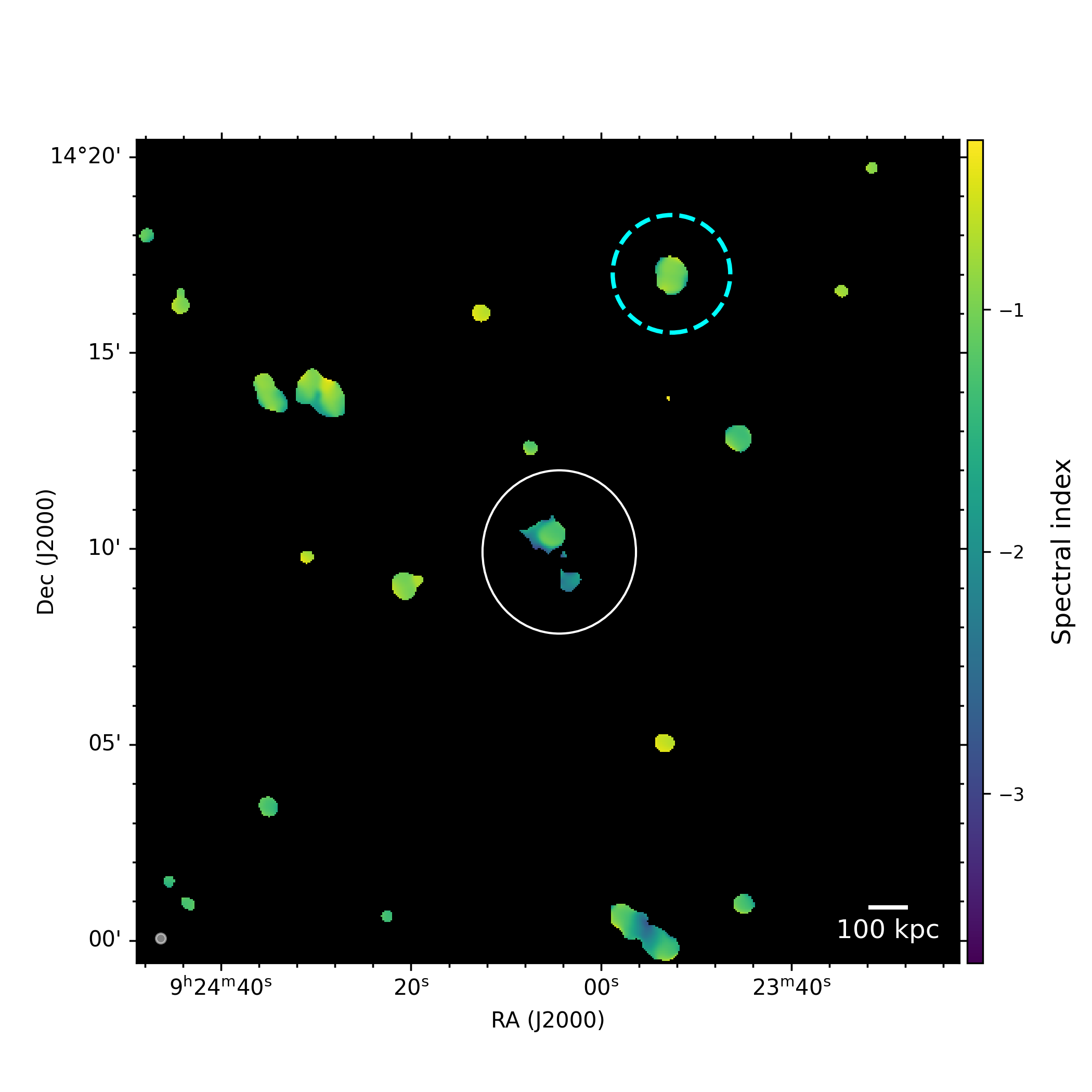}
    \caption{Wide field spectral index map of A795. The white ellipse marks the position of the extended radio emission cospatial with the BCG. The cyan dashed circle signs an additional extended X-ray source (separate from the ICM of A795), see Sec. \ref{group2D}.}
    \label{spixmap}
\end{figure}
\end{appendix}

\end{document}